\documentclass[usenames,dvipsnames,screen,format=sigconf,table,xcdraw,9pt]{acmart}
\renewcommand\footnotetextcopyrightpermission[1]{} 
\usepackage{subfigure}
\newcommand{\notes}[1]{}
\renewcommand{\notes}[1]{#1}
\usepackage[]{xcolor}

\newcommand{\eat}[1]{}
\newcommand{\ie}[0]{\textit{i.e.,}\xspace}
\newcommand{\eg}[0]{\textit{e.g.,}\xspace}

\usepackage{titlesec}
\titlespacing{\section}{0pt}{4pt}{2pt}
\titlespacing{\subsection}{0pt}{4pt}{2pt}

\usepackage{graphicx}
\usepackage[justification=RaggedRight,labelfont=bf,font=small,textfont=normalfont,tableposition=below]{caption}
\usepackage{makecell}
\usepackage{amsmath}

\usepackage{amssymb}
\usepackage{url}
\usepackage{multirow}
\usepackage[ruled,vlined]{algorithm2e}
\usepackage{enumitem}
\usepackage{listings}
\usepackage{diagbox}
\usepackage{colortbl}
\usepackage{hyperref}
\usepackage{comment}
\usepackage{balance}
\usepackage{dsfont}
\usepackage{changepage}
\usepackage{textcomp}
\usepackage{cleveref}

\setitemize{leftmargin=4mm}
\setenumerate{leftmargin=4mm}
\setlist{noitemsep,topsep=0pt,parsep=0pt,partopsep=0pt}

\newcommand{\paraskip}{\vspace{2pt}}
\newcommand{\name}{\textsf{Confucius}\xspace}

\newcommand{\elephant}{real-time\xspace}

\newcommand{\parahead}[1]{\paraskip\noindent\textbf{#1}}
\newcommand{\ithead}[1]{\paraskip\noindent\textit{#1}}

\newcommand{\remove}[1]{}

\newcommand\rurl[1]{\href{https://#1}{\nolinkurl{#1}}}

\crefname{figure}{Fig.}{Figs.}
\crefname{section}{\S}{\S}
\crefname{table}{Tab.}{Tab.}
\crefname{algorithm}{Algorithm}{Algorithms}
\crefname{equation}{Eq.}{Eqs.}
\crefrangelabelformat{section}{#3#1#4-#5#2#6}
\crefrangelabelformat{figure}{#3#1#4-#5#2#6}

\begin{document}

\title{\name: Achieving Consistent Low Latency with\\Practical Queue Management for Real-Time Communications}



\author{Zili Meng}
\affiliation{Hong Kong University of Science and Technology}
\email{zilim@ust.hk}

\author{Nirav Atre}
\affiliation{Carnegie Mellon University}
\email{natre@cs.cmu.edu}

\author{Mingwei Xu}
\affiliation{Tsinghua University}
\email{xumw@tsinghua.edu.cn}

\author{Justine Sherry}
\affiliation{Carnegie Mellon University}
\email{sherry@cs.cmu.edu}

\author{Maria Apostolaki}
\affiliation{Princeton University}
\email{apostolaki@princeton.edu}

\begin{abstract}

Real-time communication applications require consistently low latency, which is often disrupted by latency spikes caused by competing flows, especially Web traffic.
We identify the root cause of disruptions in such cases as the mismatch between the \textit{abrupt} bandwidth allocation adjustment of queue scheduling and \textit{gradual} congestion window adjustment of congestion control.
For example, when a sudden burst of new Web flows arrives, queue schedulers abruptly shift bandwidth away from the existing real-time flow(s).
The real-time flow will need several RTTs to converge to the new available bandwidth, during which severe stalls occur.
In this paper, we present \name, a practical queue management scheme designed for offering real-time traffic with consistently low latency regardless of competing flows. 
\name slows down bandwidth adjustment to match the reaction of congestion control, such that the end host can reduce the sending rate without incurring latency spikes. 
Importantly, \name does not require the collaboration of end-hosts (\eg labels on packets), nor manual parameter tuning to achieve good performance.
Extensive experiments show that \name outperforms existing practical queueing schemes by reducing the stall duration by more than 50\%, while the competing flows also fairly enjoy on-par performance. 
\end{abstract}

\maketitle

\vspace{-1em}
\section{Introduction}
\label{sec:intro}

Real-time (RT) video communications, including a range of applications from video conferencing to cloud gaming and VR/AR streaming, are becoming the dominant traffic on the Internet. 
These applications require low and consistent latency to maximize the user experience~\cite{sigcomm2022zhuge}.

Significant research has been dedicated to ensuring a satisfactory user experience through minimizing and stabilizing the end-to-end latency.
Indeed, congestion control algorithms (CCAs) reduce the queueing delay~\cite{ton2017webrtc, ray2022sqp, nsdi2018copa}; forward error correction (FEC) improves the loss recovery~\cite{nsdi2023tambur, nsdi2024hairpin}; multiple path transport mitigates fluctuation in wireless settings~\cite{mm23twinstar, sigcomm2023cellfusion, sigcomm2023converge}; while co-design with the video codec~\cite{nsdi2023afr, nsdi2018salsify} and wireless routers~\cite{sigcomm2022zhuge, mmsys2015macadapt} controls the delay in these components.
Unfortunately, these works mainly \textit{focus on how to mitigate the effect of network fluctuations after the fact, instead of addressing their root cause}. 
As a result, latency fluctuations still routinely occur, causing stalls and deterioration of the performance of the real-time flow~\cite{imc2021can, imc2022enabling}.

\begin{figure}
    \centering
    \includegraphics[width=.9\linewidth]{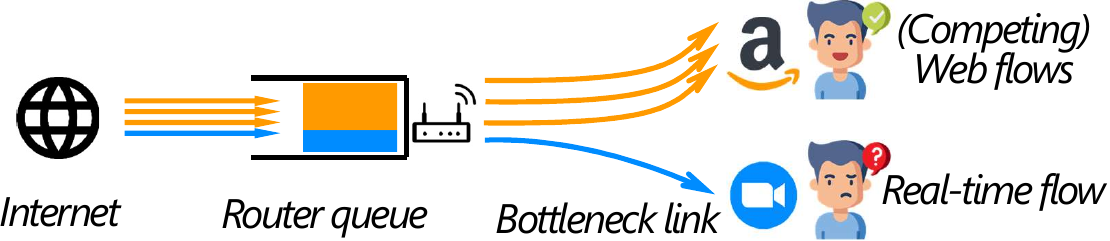}
    \caption{The scenario where the real-time flow is affected by competing flows. When Web flows join the competition with the real-time flow, the available bandwidth of the real-time flow will be immediately reduced. Note that even loading one Web page can have tens of concurrent active flows.}
    \label{fig:intro-example}
\end{figure}

In this paper, we show that unpredictable flow competition in the network layer can cause drastic network fluctuation, which drastically affects real-time flows (\cref{sec:2}).
For instance, loading a single Web page creates nine concurrent Internet connections (on average), drastically reducing the available bandwidth for the competing real-time flows and causing stalls in multiple practical settings such as home routers, as shown in \cref{fig:intro-example}. 
Congestion control alleviates the issue by reducing the real-time flow's congestion window or sending rate after end hosts observe latency increases or packet loss, but it is already far too late.
Indeed, it will take several RTTs for congestion control to react and converge to the new available bandwidth, while the \textit{packets sent in excess of the allocated bandwidth} during the convergence period will lead to an increase in the end-to-end delay. 
These endpoint-based optimizations, in general, cannot fundamentally prevent such performance degradation from happening since the onset of competing traffic is unpredictable. 

A natural solution to flow competition is to manage the router queue and prevent the available bandwidth of the real-time flow from reducing.
There have been works trying to achieve this for several decades.
In differentiated services (DiffServ)~\cite{rfc4594diffserv} (including L4S~\cite{rfc9330l4s}), the router recognizes the pre-defined labels (priorities) from packet headers and schedules packets based on these labels.
However, such a design is not incentive-compatible in practice: applications have the incentive to mark their packets with higher priority, which eventually leads to the Tragedy of Commons -- routers will not respect the labels, and endpoints cannot count on using them.
Another category of solutions on the router is active queue management (AQM), which tries to notify the sender in advance before the queue builds up~\cite{cacm2012codel, sigcomm1997red, sigcomm2022zhuge}.
We demonstrate in \cref{sec:2} that these are still reactive mechanisms and cannot prevent stalls from happening.

We argue that the root cause for stalls in the real-time flow is the \textit{mismatch in reaction time between the bandwidth allocation mechanisms on routers and rate adaptation mechanisms on endpoints}.
In \cref{fig:intro-example}, when nine new flows triggered by a single website (in yellow) suddenly compete with an existing real-time flow (in blue), the available bandwidth of the real-time flow is \emph{immediately} reduced to 1/10 of what it was. 
However, the sender's congestion window needs several round-trip times (RTTs) to gradually adjust to match the new available bandwidth. 
During this adjustment period, packets that are sent in excess of the allocated bandwidth will induce congestion, resulting in bufferbloat and stalls.
While the Web flows will complete within one or two seconds and relinquish their bandwidth share, the real-time flow will have already experienced significant degradation.
We find that existing queue scheduling and management algorithms \textit{ignore the transient temporal behaviors during the network change}, leading to the stalls.
This highlights a critical need for a queue management scheme that takes into account the convergence time of the congestion control to prevent such stalls.

To this end, we designed \name,\footnote{
    \name' (the philosopher) educational philosophy is teaching students by their essences. In this paper, we serve the flows by their essences.
} a practical queue management scheme that aims at providing consistent low latency for real-time flows independently of competing flows at the bottleneck. Instead of abruptly changing bandwidth allocations when a burst of new flows arrive, \name{} {\it gradually} adjusts the service rates to provide existing flows a few RTTs to detect the change in network conditions and adjust their congestion windows. 
In this case, the excessively sent packets will be reduced and the latency for the real-time flow can be maintained.

We design \name{} to fulfill three fundamental requirements related to consistency, fairness and incentive-compat-ibility (\cref{sec:design-req}):
First, \name{} needs to provide latency consistency to real-time flows independently of the number, rate, or congestion control of the competing flows. \name{} achieves this by offering a theoretical upper bound for  latency fluctuation experienced by real-time flows, which we also validate through experiments
Second, \name should eventually be fair.
For instance, in \cref{fig:intro-example}, the performance of Web flows should not be sacrificed. To achieve this, \name{} smoothly moves service rates towards the fair allocation within a few RTTs.
Finally, \name's classification of real-time flow should be practical and on-router, without relying on end hosts for traffic classification. Unfortunately, the alternative -- flow classification algorithms -- are usually expensive and sensitive to protocols~\cite{hotnets2019inc}.
\name{} calssifies flows by \textit{how aggressively they occupy the buffer} at the bottleneck router, a metric that directly reflects how important low latency is to a flow.

We implement \name{} with both NS-3 simulator and kernel modules on Linux-based routers.
Note that \name is designed for last-mile routers (e.g., home routers in \cref{fig:intro-example}) where the competition can lead to congestion, and the computation is more flexible since home routers are mainly Linux-based\footnote{
    A measurement shows that 91\% of home routers are Linux-based~\cite{HomeRouterSecurity}. 
}.
With real-world bandwidth and Web page traces, we show that compared to FqCoDel (Linux's default)~\cite{rfc8290fqcodel}, \name reduces the stall duration of the real-time flow by 60\%-69\% and the loading time of Web pages by 39\%-48\% for top 1000 websites at the same time.
Compared to other baselines that do not require labels from end hosts, \name can still effectively reduce the stall duration of the real-time flows by at least 21\% (\cref{sec:eva-trace}) with negligible computation overhead (\cref{sec:eva-testbed}).
In the meantime, long-lived, bulk transfers experience no degradation at all relative to fair queueing, and the impact on the flow completion time over short Web flows is limited to at most 10\% even compared with the shortest job first (\texttt{SJF}, which strictly prioritizes the Web flows).
We will release all traces and codes of this paper.

\section{Motivation}
\label{sec:2}

We start by describing recent trends that call for consistent low latency (\cref{sec:motiv-back}). 
Next, we explain via an intuitive example why existing solutions fail to achieve consistent low latency under flow competition (\cref{sec:motiv-example,sec:motiv-cause,sec:motiv-related}).

\subsection{The rise of real-time traffic} 
\label{sec:motiv-back}

While the Internet has always been shared among multiple applications, the proliferation of real-time communication applications (\eg{} videoconferencing, cloud gaming, virtual reality) has made sharing of bottleneck links particularly challenging. 
Real-time applications require not just low latency but \emph{consistently} low latency while sending at moderate to high throughputs (ranging from tens to hundreds of Mbps)~\cite{sigcomm2022zhuge, sigcomm2020tack, imc2022measurement}. 
For real-time applications, latency consistency is extremely critical to user experiences.
For example, a transient increase in latency to 200 ms might cause cloud gaming users to lose~\cite{qomex2015cgsteam}.
Therefore, controlling the latency fluctuation and achieving a consistent low latency for real-time applications is essential.

\parahead{Setting \& Scope:} 
This work focuses on end-user access points (\eg{} wireless or wired home routers), where it is well-known that congestion and latency fluctuation are frequent~\cite{sigcomm2022zhuge, ccr2017lastmile, imc2020persistent}.
Despite recent advances in wireless technologies such as 5G and WiFi 6, the last-mile access routers are still likely to be the cause of jitter, irrespective of whether the last-mile is wired~\cite{nsdi2023crab} or wireless~\cite{sigcomm2020measure5g, imc2017fastack, sigcomm2022zhuge}. 
As most such routers are Linux-based~\cite{linux-router, HomeRouterSecurity}, they allow for flexible traffic management on software which is a great opportunity for innovation. 
Our experiments and data involve applications used in those settings. 
We also include a benchmark in a Linux-based router \cref{sec:eva-testbed}).
Congestion in other settings (\eg{} losses in the Internet core~\cite{sigcomm2018inferring} or datacenters~\cite{nsdi2015pias}) are out of scope for this work. 


\subsection{Motivating example}
\label{sec:motiv-example}

\begin{figure}
    \includegraphics[width=\linewidth]{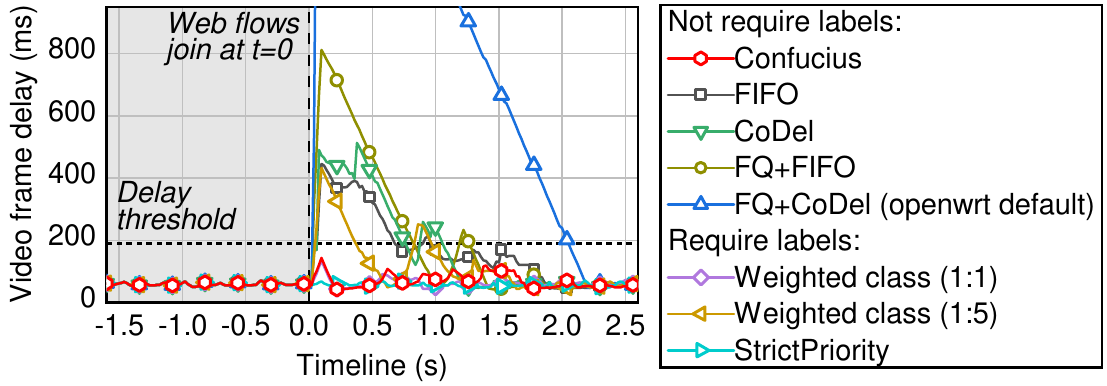}
    \caption{An existing real-time flow competes with flows of loading the homepage of\rurl{amazon.com}, as shown in \cref{fig:intro-example}. The real-time flow, using GCC~\cite{ton2017webrtc}, always experiences transient stalls during the competition unless flows are pre-labeled by the end host and differentiated by the router.}
    \label{fig:motiv-timeline}
\end{figure}

To better illustrate the problem and the limitations of existing approaches, we revisit the example of \cref{fig:intro-example} in detail.
Consider a user who is on a video call, and their housemate (with whom they share the home router) decides to load a Web page. 
Technically, one existing real-time flow on the bottleneck will compete with the new flows from one Web page. 
We simulate the real-time flow's delay of each video frame using NS-3 and present the results in \cref{fig:motiv-timeline} (details in \cref{sec:eva-impl}).
Before considering other queue management mechanisms, let us focus on the performance of \texttt{FIFO} (square markers).

Before t=0s, the sending rate of the real-time flow has converged, with the video frame delay fluctuating around 60ms, which is much lower than the stall threshold (190ms\footnote{
    This is the recommended network delay for video chats by ITU~\cite{itu-ddl}.
}) required by the application.
However, when flows from loading the homepage of\rurl{amazon.com} join the competition on the bottleneck, the end-to-end delay for the real-time flow sharply increases.
Using \texttt{FIFO}, the delay goes up to more than 400 ms, and stays above the threshold for almost one second, during which a stall occurs and the user experience is impaired.
When using \texttt{FqCoDel}, the delay of the real-time flow is even worse since fair queueing shifts more bandwidth away and \texttt{CoDel} drops more packets.
We find that the delay \textit{always} spikes regardless of the underlying CCA (\cref{sec:eva-trace}).

\subsection{Root cause analysis}
\label{sec:motiv-cause}

We argue that the delay spike is caused by (i) the burst of flows and packets from the competing Web page; (ii) the abrupt reallocation of the available bandwidth by queue management; and (iii) the gradual reaction from the congestion control.
Next, we will elaborate on how these common factors result in performance degradation, and explain the limitations of existing works.

\begin{figure}
    \centering
    \subfigure[Timeline for\url{amazon.com}.]{
        \includegraphics[height=2.6cm]{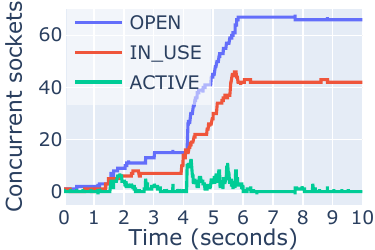}
        \label{fig:web-amazon}
    }
    \subfigure[Distributions.]{
        \includegraphics[height=2.6cm]{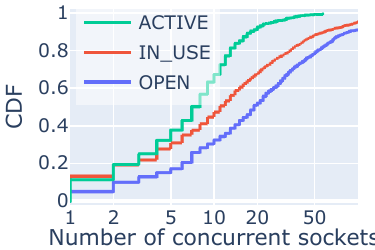}
        \label{fig:web-socket-cdf}
    }
    \caption{Number of \textit{concurrent} flows recorded by \texttt{NetLog}~\cite{netlog}. \texttt{OPEN} and \texttt{IN\_USE} are socket states marked by Chrome, and \texttt{ACTIVE} means that the flow is receiving bytes in the last 10 ms.}
    \label{fig:web-measure2}
\end{figure}

\parahead{The source of the burst: One Web page triggers multiple, concurrently-active flows.}
To understand the burst in \cref{fig:motiv-timeline}, we measure the flows triggered by\rurl{amazon.com} over time.
Concretely, we measure the number of sockets that are \texttt{OPEN} and \texttt{IN\_USE} marked by NetLog~\cite{netlog} from Chrome.
We also measure the number of active flows that receive bytes every 10 ms through packet captures (\texttt{ACTIVE}). 
As shown in \cref{fig:web-amazon}, loading only the homepage generates up to 68 flows in total, where up to 12 flows run simultaneously.
This is due to the Web design of hosting different objects (\eg images, videos, ads, scripts) in various domains.
Note that this is not due to the parallel connections in HTTP/1.1 -- we later present in Appx.~\ref{app:web} showing that more than half of the flows go to different unique IPs.

This triggering of multiple flows to load one page is shared across different websites.
We measured the homepage of Top 1000 websites in November 2023 from the saved Alexa list and presented the distribution in \cref{fig:web-socket-cdf}.
We find that the median number of concurrent \texttt{ACTIVE} flows is 8 while the 90th percentile is 19.
The highest one in the Top 200,\rurl{dailymail.co.uk}, has up to 50 active flows and 250 open sockets at the same time.
We present the structure and list some famous websites in Appx.~\ref{app:web}.
Moreover, for some websites (\eg{} Wikipedia and Google), loading other pages triggers more flows compared to the almost blank home page, which will further exacerbate the degradation experienced by the real-time flow.

\begin{figure}
\centering
\vspace{-.5em}

\subfigure[FIFO.]{
    \includegraphics[width=3.8cm]{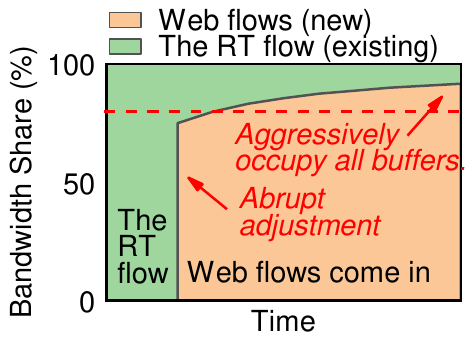}
    \label{fig:design-weight-fifo}
}
\subfigure[FQ.]{
    \includegraphics[width=3.8cm]{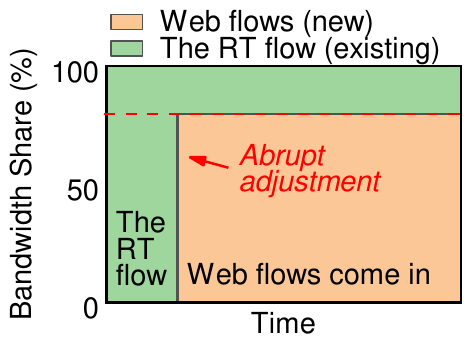}
    \label{fig:design-weight-fq}
}

\subfigure[Weighted class (1:1).]{
    \includegraphics[width=3.8cm]{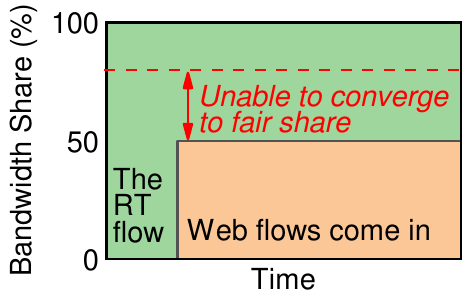}
    \label{fig:design-weight-drr}
}
\subfigure[\name.]{
    \includegraphics[width=3.8cm]{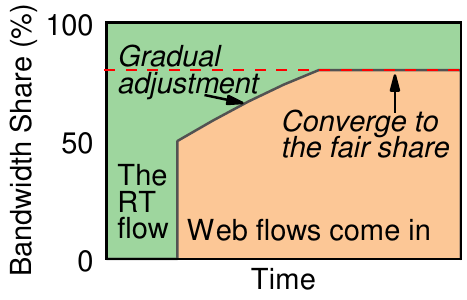}
    \label{fig:design-weight-name}
}
\caption{Illustration of how bandwidth shares change over time with incoming Web flows and the existing real-time (RT) flow for different schedulers. The dashed red line marks the fair share.}
\label{fig:design-weight}
\end{figure}

\parahead{The cause of the delay spike: Queue schedulers sharply reallocate service rates.} 
Queue management typically reacts to the instant conditions of all flows in the queue.
Revisiting our example, when the page loading starts, tens of packets of Web flows immediately arrive at the bottleneck, creating a queue.
At the same time, the real-time flow only has a few packets in the queue since it always tries to keep the queue near-empty~\cite{nsdi2018copa, ton2017webrtc}.
We illustrate the bandwidth share of different queue management schemes in \cref{fig:design-weight}.
For \texttt{FIFO} (\cref{fig:design-weight-fifo}), the service rates for different flows are proportional to the number of bytes per flow in the queue, thus, the available bandwidth for the real-time flow will be drastically reduced.
Fair queueing (\texttt{FQ}, \cref{fig:design-weight-fq}) makes matters worse and allocates even less bandwidth to the real-time flow, since those short Web flows are many more than the real-time flow.
Concretely, in the\rurl{amazon.com} example, 12 new flows joining the fair queueing router will directly reduce the available bandwidth of the real-time flow to 1/13.

\begin{figure}
    \centering
    \includegraphics[height=2.5cm]{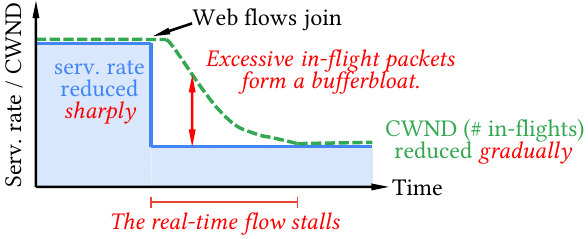}
    \caption{When new competing flows join, the service rate of the real-time flow will be immediately reduced, but the CCA takes multiple RTTs to converge.}
    \label{fig:rate-dec}
\end{figure}

Such a sharp decrease in the available bandwidth causes a delay spike to the real-time flow.
This is because the CCA needs to \textit{gradually} probe and match its sending rate to the new available bandwidth, which takes several RTTs (dashed green line in \cref{fig:rate-dec}).
While the number of in-flight packets is converging to the new bandwidth-delay product, the excessive in-flight packets will cause bufferbloat and result in high end-to-end latency for the real-time flow.

Active queue management (AQM) algorithms, which notify the sender about the network conditions by dropping or marking ECN on packets, cannot prevent stalls either.
This is mainly because flows driven by different congestion control algorithms (CCAs) have different perceptions of congestion (\eg{} delay, loss, rate). 
Therefore, as shown in \cref{fig:motiv-timeline}, CoDel~\cite{cacm2012codel} leads to a latency spike even higher than that of \texttt{FIFO}. 
We observe similar limitations in other AQMs (\cref{sec:eva-trace}).

\parahead{The fact hard to change: Congestion control takes a longer time to converge.}
As we discussed, the issue is when the competing flows join, the available bandwidth for the real-time flow drops immediately, but the end-to-end CCA cannot immediately reduce the inflight packets to fit the new available bandwidth. 
End-to-end CCAs do not know how much to reduce and have to reduce step-by-step.

Some proposals are designed to help the CCA to quickly converge to the new available bandwidth, such as XCP~\cite{sigcomm2002xcp}, RCP~\cite{infocom2008rcp}, Kickass~\cite{icnp2016kickass}, and ABC~\cite{nsdi2020abc}.
However, none of these proposals work unless both end hosts and routers collaboratively deploy these protocols and offer no improvement otherwise.
This poses significant barriers to deployment on the Internet~\cite{sigcomm2022zhuge}.
Moreover, during the convergence of the CCA, the excessive in-flight packets also inflate the RTT.
For most CCAs using RTT to update (\eg{} adjust the sending rate every RTT), the update period will, in turn, inflate after the first several packets.
In the example in \cref{fig:motiv-timeline}, before the Web flows join, the RTT for the real-time flow is around 40 ms.
However, during the competition, the RTT inflates to hundreds of milliseconds.
Putting all factors together, 
we can see that for all baselines that do not require labels, the delay spike of the real-time flow goes up to at least 400 ms.

\subsection{Limitations of related works}
\label{sec:motiv-related}

One line of solution is DiffServ~\cite{rfc4594diffserv}, which labels the flows of interest in advance and schedules them differently on the router using \texttt{StrictPriority} or \texttt{weighted class} as shown in \cref{fig:motiv-timeline}.
This also includes the recent proposal L4S~\cite{rfc9330l4s} which we will later evaluate in \cref{sec:eva-trace}.
While this is deployable in datacenters~\cite{sigcomm2016karuna}, it is not practical on the Internet.
End hosts have the incentive to fake their labels if that could help their flows have better performance.
It is also challenging to coordinate the end host and router on the Internet in the real world since they usually belong to different entities.
Even with perfect labels, achieving optimal performance requires optimal allocation of bandwidth across the different classes of traffic. 
To understand why this is challenging, consider some canonical solutions.
\texttt{StrictPriority}, albeit guaranteeing the latency for the real-time flow, will drastically harm the performance of competing Web flows (\cref{sec:eva-trace}).
Allocating bandwidth for different classes using pre-defined weights needs accurate estimation of the bandwidth demands from both classes, where inaccurate estimation easily leads to unfairness or latency spikes.
For example, if we set the ratio between the real-time flows and Web flows to 1:1, the Web flows will suffer from degraded PLT since they cannot obtain their fair shares (\cref{fig:design-weight-drr}), while 1:5 will lead to the latency spike to the real-time flow as well (\cref{fig:motiv-timeline}).

There are further mechanisms as below, which, unfortunately, still reactively respond to network changes. 
Zhuge~\cite{sigcomm2022zhuge} reduces the feedback loop between the router and the endpoint from one RTT to sub-RTT levels, but CCA convergence still requires multiple RTTs (\cref{sec:motiv-cause}).
Using the example in \cref{fig:rate-dec}, Zhuge tightens the turning point of the green dashed line, but the dominant contributor to delay -- the time it takes for the green dashed line to converge to the blue line -- persists.
FEC is designed for loss recovery~\cite{nsdi2023tambur, nsdi2024hairpin} and is hardly helpful in our example since most of them have no loss at all.
Multipath transport will switch to the new path~\cite{sigcomm2023converge, sigcomm2023cellfusion, nsdi2024augur}, but this also occurs \textit{after} the sender observes drastic degradation in the current path.
Real-time flows still have to suffer from stalls during the adaptation period. 
Bandwidth estimations from the wireless link layer and below~\cite{sigcomm2022zhuge, sigcomm2020pbecc} are not effective either since the link capacity does not change in the competition.

\eat{
\section{Taming Volatility While Achieving Fairness}
\label{sec:insight}

As described in \S\ref{sec:intro}, there exists a \textit{fundamental} tension between fairness and non-volatility (\eg consistent latency for jitter-sensitive video-conferencing traffic).
Consider again our motivating example of an HRT flow that begins competing with a burst of \textit{N} newly joining short flows on a bottleneck link. On the one hand, we could `shield' the HRT flow from volatility by always giving it a higher share of the bandwidth.
However, this unfairly penalizes short flows, grossly violating the fairness objective (\eg CBQ will degrade the PLT for Web flows by more than 100\% sometimes compared to FQ, as later shown in Fig.~\ref{fig:cdf-copa-plt} in \S\ref{sec:eva-trace}). Conversely, we could enforce strict fairness by immediately re-partitioning the bandwidth equitably between the different flows (\ie reducing the \elephant flow's bandwidth limit to $\frac{1}{N+1}$ of the link capacity); however, in the time it takes the CCA's control loop to detect and lower its sending rate to match its allocated bandwidth (\textit{at least} one RTT, but may be dozens~\cite{sigcomm2022zhuge}!), the damage is done.
Especially when this happens at the last-mile routers where buffers are deep~\cite{sigcomm2022zhuge}, packets sent during this period will experience significant queueing delay.

Looking closely, the performance degradation has two causes:

\begin{enumerate}[label=(\roman*), leftmargin=6mm]
    \item The sender needs some time to detect congestion and reduce its sending rate. In this case, the responsiveness of the CCA and tightness of its feedback loop will affect the degradation -- the less responsive the CCA is, the longer it takes to reduce the sending rate. 
    \item Packets sent in excess of the available bandwidth during this time will end up being queued at the bottleneck queue, and must wait to be transmitted. In this case, the magnitude of the available bandwidth drops matters -- the more substantial the available bandwidth drop, the longer it takes to drain the queue. Even if one ideal CCA can reduce its sending rate to zero immediately after detecting the changes in the bottleneck, the inflight packets sent during the feedback loop still need considerable time to drain when the magnitude of available bandwidth drop is high.
\end{enumerate}

Putting these together, we observe that the tension between fairness and non-volatility is an artifact of the disparity between the \elephant flow's instantaneous sending rate (as determined by the underlying CCA's control loop) and the available bandwidth during the transient period (as determined by a strictly-fair bandwidth allocation).
}
\section{\name Design}
\label{sec:design}

Our previous observations motivate \name, a practical queue management scheme for achieving consistent and low latency for real-time flows that is designed to work on home routers. 
We describe \name's design requirements in \cref{sec:design-req} before we give an overview of \name on \cref{sec:design-overview}.

\subsection{Design Requirements}
\label{sec:design-req}

\vspace{-2pt}
\parahead{R1: The performance of the real-time flow should be robust to any competing flows.}
\name{} stands out among queue management algorithms in that it theoretically guarantees worst-case performance, no matter what congestion control algorithms and competing flows are.
This will, in consequence, fundamentally address the root cause of latency fluctuation induced by unpredictable competing traffic.
It is easy to vaguely describe \name{} as `controlling latency fluctuations' but it is harder to formulate this into a rigorous service model. 
We theoretically calculate performance bounds for a few classes of applications that might use \name{}. 
We demonstrate that with \name{}, real-time flows have a near-constant bound of latency degradation (around 250 ms in \cref{sec:eva-workload}), no matter how large and how many competing flows join the bottleneck.

\parahead{R2: Latency consistency should not come at the cost of long-term fairness.} 
\name{} should still follow per-flow fairness in the long run. 
To do so, \name{} moves rates towards a fair allocation quickly and pushes the blue solid line in \cref{fig:rate-dec} to match the green dashed line.
In this case, the latency spike will be controlled and the bandwidth for the competing flows will be largely protected as well.
Technically, \name{} adjusts the service rate of flows using exponentially weighted moving average (EWMA)~\cite{lucas1990ewma}, as shown in \cref{fig:design-weight-name}.
This allows the CCA to gradually react following the bandwidth share of \name{}, and also converges to the fair share in several RTTs.
Note that the RTT is not inflated due to the excessive packets.
Our experiments (\cref{sec:eva-workload}) and theoretical analysis (\cref{sec:weight-theory}) show that such a design can effectively achieve fairness and latency consistency.

\parahead{R3: The identification of real-time flows should not rely on end hosts.} 
A naive solution is to split the flows by their age. 
However, this is not practical since flows driven by different CCAs or having distinct objectives should not share the same queue either.
Meanwhile, using FQ to split old flows cannot provide low latency to the bursty flows~\cite{macgregor2000deficits}, which is usually the case for real-time video streaming.
Thus, we still need to identify different types of flows.
To make \name{} incentive-compatible and deployable in practice, we aim to identify the flows of interest \textit{at the router itself}, without relying on end hosts.
The performance improvements should be directly observed by the router vendor without going through endless coordination between end-host content providers and router vendors in IETF.
In \cref{sec:design-classify}, we illustrate how \name{} identifies flows based on their queue occupancy: built on the CCA evolution, real-time flows naturally occupy a small fraction of the buffer (e.g., GCC~\cite{ton2017webrtc, nsdi2018copa}), while throughput-oriented flows are observed to be buffer-filling (e.g., Cubic).
\name uses the queue occupancy to differentiate the flows in the queue.

\subsection{Design Overview}
\label{sec:design-overview}

\begin{figure}
    \centering
    \includegraphics[width=\linewidth]{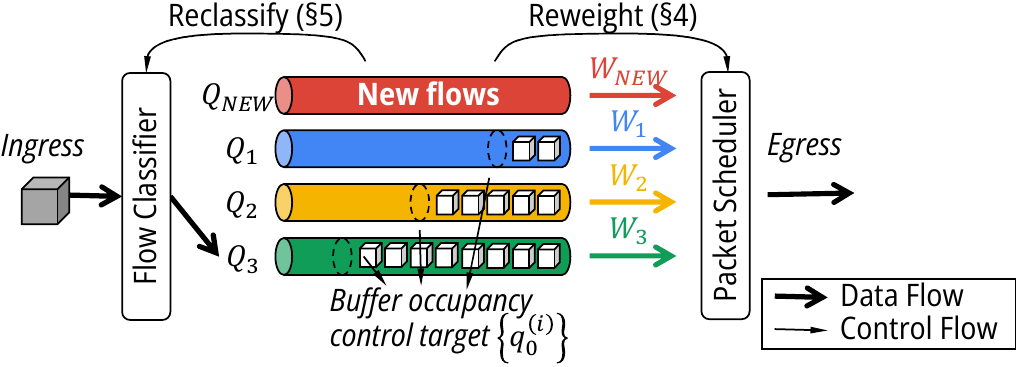}
    \caption{Design overview of \name. $w_i$ denotes the weight for queue $i$ in the scheduling with DWRR.}
    \label{fig:overview}
\end{figure}

At a high level, \name classifies flows to queues and strategically assigns a portion of the link capacity to each of them, as illustrated in \cref{fig:overview}.

To address the goal \textbf{R1} and \textbf{R2}, \name leverages a simple yet powerful insight from \cref{sec:design-req}: 
Upon the arrival of competitors, the reduction of the available bandwidth of existing flows is inevitable if we want to preserve long-term throughput fairness.
Yet, we can gradually and cautiously control the reduction of the available bandwidth during the transient period.
Thus, we can eliminate the mismatch between the sending rate of the CCA and the service rate at the bottleneck link for existing real-time flows, thereby taming the latency fluctuation. 
We will extend our insight of using the EWMA reweight mechanism in \cref{sec:design-weight}.

For \textbf{R3}, by grouping flows with similar queue occupancy into the same queue, flows with different queue occupancies will not affect each other. 
Meanwhile, with a fixed number of queues to schedule between (instead of per-flow queues such as \texttt{FQ}), latency-sensitive flows will have a consistent latency.
Thus, \name uses a set of queues ($Q_1,Q_2,Q_3$), each designed to accommodate old flows with different buffer occupancies, and a separate queue ($Q_{NEW}$) dedicated to new flows. 
It then adopts a Deficit-Weighted Round-Robin (DWRR) algorithm to schedule between these queues. 
Finally, \name periodically measures flow characteristics and reclassifies flows using a hysteresis-based mechanism to further increase robustness in practice (\cref{sec:design-classify}).

\section{Age-aware Flow Weights Adjustment}
\label{sec:design-weight}

In this section, we explain the benefits of exponential bandwidth reallocation (\cref{sec:weight-overview}) and dive into \name' weight adjustment (\cref{sec:weight-mechanism}).
We then analytically show that it guarantees bounded performance degradation, both for existing \elephant flows and newly-arrived competing flows (\cref{sec:weight-theory}).

\subsection{Exponential bandwidth re-allocation}
\label{sec:weight-overview}

\begin{figure}
    \centering
    \subfigure[Measurements]{
        \includegraphics[height=2.6cm]{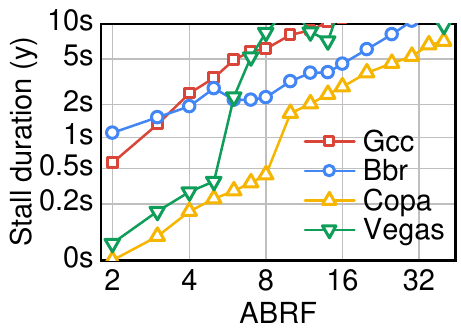}
        \label{fig:jitter-reduc-exp}
    }
    \subfigure[Illustration]{
        \includegraphics[height=2.6cm]{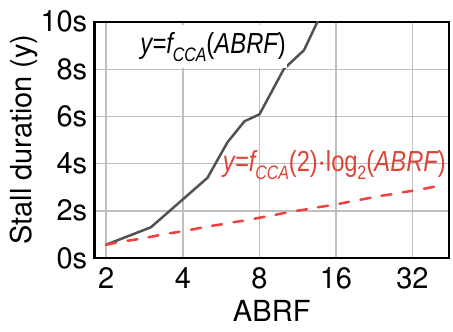}
        \label{fig:jitter-reduc-illu}
    }
    \caption{(a) Stall duration increases with the available-bandwidth-reduction factor (ABRF). (b) An illustration of how gently reducing available bandwidth helps reduce delay duration. Note that (a) is a log-log plot but (b) is a log-lin plot.}
    \label{fig:jitter-reduc}
\end{figure}

We first quantitatively demonstrate the advantage of \emph{gradually} controlling the \elephant flow's bandwidth allocation compared to directly cutting its available bandwidth to its fair share.
We measured the stall duration $y$ for the real-time flow in the scenario of a sudden reduction of available bandwidth for four low-latency CCAs (\cref{sec:design-weight}). 
Concretely, $y$ denotes the stall duration defined by more than 190 ms of end-to-end delay. 
We plot $y$ as a function of the \textit{Available Bandwidth Reduction Factor} (ABRF, the factor we will reduce the available bandwidth) for different CCAs in \cref{fig:jitter-reduc-exp}. 
We find that CCAs respond poorly to sudden, large reductions in bandwidth.
For instance, reducing GCC's available bandwidth to 1/16 of its initial value (i.e., $ABRF=16$) results in a \textit{$y > 10$ seconds} stall. 
The relationship between the stall duration and ABRF ($y=f_{CCA}(ABRF)$) is super-linear.

To avoid such stalls, \name \textit{gradually} reduces the available bandwidth for the real-time flow.
For instance, to achieve a final ABRF of 16, we can reduce the available bandwidth four times, each by half.
\cref{fig:jitter-reduc-illu} demonstrates, in the ideal case, the value proposition of this approach.
Compared to the super-linear stall duration (solid line copied from \cref{fig:jitter-reduc-exp}), exponentially reducing the sending rate will only increase the stall duration \textit{logarithmically} with the ABRF (modulated by $f_{CCA}(2)$, a small constant).

Such a smooth reallocation of available bandwidth allows the CCA to learn the reduced bandwidth allocation, and is also robust to the number or size of competing flows.
No matter how many flows compete with the real-time flow, the curve of the available bandwidth of the real-time flow is fixed so the delay will remain the same.
Meanwhile, adjusting the bandwidth share exponentially yields fast convergence to the fair share, satisfying requirements \textbf{R1} and \textbf{R2} together.
We prove in \cref{sec:weight-theory}, that \name guarantees that the long-term fairness will not be impaired, and the degradation of the performance for new flows will \textit{always be within a constant, additive factor} of the FCT under a strictly fair allocation.

\subsection{Adjustment Mechanism}
\label{sec:weight-mechanism}

To assign service rates to queues, \name uses the following process.  
For each flow, $f$, \name computes a weight, $w_f$, to represent its share of the bandwidth (service rate).
\name groups new flows into a separate queue called $Q_{new}$ (depicted in \cref{fig:overview}). 
All existing flows which are mapped to other queues are assigned flow weights of $w_f=1$, and are collectively denoted as set $\mathcal{F}_{ext}$. 
The flow weights of all flows in $Q_{new}$ are computed as follows:
\begin{equation}
    \small
    \label{eq:flow-weight}
    w_f = \min\left(\frac{|\mathcal{F}_{ext}|}{|Q_{new}|}\cdot 2^{\lambda t},\ 1\right),\quad f\in Q_{new}
\end{equation}
Then, for a given queue, $Q$, the weight is the sum of weights of all flows in $Q$.
There are several considerations in \cref{eq:flow-weight}:

\parahead{Age-aware exponential adjustment $\left(2^{\lambda t}\right)$.} 
As described in \cref{sec:design-req}, \name \textit{exponentially increases} the weights of new flows, where the bandwidth shares are illustrated in \cref{fig:design-weight-name}.
Here, $t$ represents the age (in milliseconds) of the new flow, and $\lambda$ is a parameter that controls the speed for the rate adjustment of new flows -- their flow weights double every $\frac{1}{\lambda}$ milliseconds.
A large $\lambda$ (\eg $\lambda\to\infty$) leads to abrupt reductions in available bandwidth and causes latency spike, while a small $\lambda$ (\eg $\lambda= 0$) results in unfairness for new flows.
Consequently, we configure $\lambda$ so that the available bandwidth for the real-time flow drops as fast as possible but not overtaking the responsiveness of the underlying CCAs.

Moreover, different CCAs have different response times to congestion.
For example, Copa needs 5 RTTs to reduce its sending rate, while BBR's response time is dictated by its probing interval of 6-8 RTTs.
To deal with the heterogeneity of CCAs on the Internet~\cite{sigmetrics2020gordon}, we set $\lambda$ as the inverse of \textit{the response time of the least responsive CCA} among common latency-sensitive CCAs.
This ensures that even the least responsive CCA can smoothly react to bandwidth changes.
Recall that we measure how different CCAs respond to bandwidth reductions in \cref{fig:jitter-reduc-exp}, which shows BBR being the least responsive CCA:
When the ABRF is 2, BBR suffers from the longest stall compared with other CCAs due to its a long probing period of 6-8 RTTs.
Thus, given a typical RTT of 30-50 ms for Web services~\cite{www2021wisetrans}, we set $\lambda$=0.004 (ms$^{-1}$) to have a doubling interval of $\frac{1}{\lambda}$=250 ms, matching BBR's probing period.
Experiments in \cref{sec:eva-trace} demonstrate satisfactory results for not only BBR but also other CCAs.


\parahead{Initial weight $\left(\frac{|\mathcal{F}_{ext}|}{|Q_{new}|}\right)$.}
To allocate sufficient share for new flows, we scale the initial weight of new flows with the \textit{number of existing flows}.
For each new flow, we set the initial weight to $\frac{|\mathcal{F}_{ext}|}{|Q_{new}|}$, where $|\mathcal{F}_{ext}|$ and $|Q_{new}|$ are the numbers of existing and new flows, respectively.
This can limit the bandwidth reduction for existing flows to be less aggressive than a factor-of-2 reduction.
In this case, the stall duration can logarithmically scale from $f_{CCA}(2)$, as shown in \cref{fig:jitter-reduc-illu}.

\parahead{Upper bound $\left(\min (...,\ 1)\right)$.} 
\name uses a flow weight threshold of $1$ to `age out' new flows from the $Q_{new}$ queue.
Once the flow weight of a flow reaches 1, the flow is no longer considered new and is moved to one of the other queues based on the output of the Flow Classifier (\cref{sec:design-classify}).



\subsection{Theoretical Analysis}
\label{sec:weight-theory}

We still follow the same example in \cref{sec:motiv-example}.
Consider one \elephant flow running by itself on a bottleneck link.
At $t=0$, $N$ new flows, each with size $B$, join the same bottleneck link and compete with the existing flow.
$B_0$ is the initial congestion window for Web flows.
We show that \name guarantees bounded stall for the existing real-time flow while yielding FCTs for Web flows within a constant additive factor of what FQ provides.
For simplicity, we summarize the results in \cref{tab:theory-summary} and leave the analytical details to Appx.~\ref{app:theory}.

For FQ and FIFO, we observe that the stall duration ($q^{max}_{P}$) scales linearly with the number of new flows, $N$, and is therefore unbounded, where $N$ can go to more than 100 in some Web pages (\cref{fig:web-measure2}).
This is quite straightforward -- when $N$ flows start to compete with the real-time flow, the available bandwidth of the real-time flow drops to $1/N$.
Intuitively, as $N$ increases, the more the available bandwidth for the \elephant flow drops, resulting in drastic delay fluctuation.

\begin{table}
    \centering
    \small
    \begin{tabular}{ccc}
    \hline
    Policy $P$ & $q^{max}_{P}$  & $T_{P}-T_{FQ}$ \\
    \hline
    FQ & $\approx{\color{red} \textit{\textbf{N}}}\left(\frac{2}{3}\sqrt{\frac{2}{k}}+q_0+\tau\right)$ & 0 \\
    FIFO & $\approx\left(\frac{{\color{red} \textit{\textbf{N}}}B_0}{q_0C}+1\right)\left(\frac{2}{3}\sqrt{\frac{2}{k}}+q_0+\tau\right)$ & $\lessapprox 0$\\
    CBQ & $\approx\frac{2}{3}\sqrt{\frac{2}{k}}+q_0+\tau$ & $\approx\frac{({\color{red} \textit{\textbf{N}}}-1){\color{red} \textit{\textbf{B}}}}{C}$ \\
    \name & $\approx 6q_0 + 15\tau + \frac{8\lambda}{k} + \frac{(10q_0 + 15\tau)\lambda^2}{k}$ & $\approx\frac{\log_2 e}{\lambda}$\\
    \hline
    \end{tabular}
    \caption{Approximations for different schedulers $P$ on their maximum queueing delay ($q^{max}_{P}$) and FCT degradation against \texttt{FQ} ($T_{P}-T_{FQ}$). \name has a bounded performance degradation for all flows. In the competition, existing schedulers have either unbounded delay, or unbounded FCT degradation. The unbounded terms with workload changes ($N$ and $B$) are marked in red.}
    \label{tab:theory-summary}
\end{table}

For class-based queues (CBQ, weighted class), pre-labeling the \elephant flow enables the scheduler to allocate the real-time flow with a fixed bandwidth share, resulting in a constant stall.
However, if the weights are not accurate (i.e., not matching the traffic ratio), CBQ converges unfairly, and the FCT degradation for new flows becomes unbounded (\cref{sec:motiv-related}).

Finally, \name yields bounded performance degradation for \textit{both sets of flows}. 
On one hand, \name ensures that the stall for \elephant flows is constant only depending on the CCA's latency sensitivity (denoted by $q_0$), the responsiveness ($k$), the feedback loop ($\tau$), and \name's decay parameter ($\lambda$)\footnote{
    When using Copa with an RTT of 40ms, $q_{\name}^{max}$ is $\approx$640 ms.
    As we show experimentally in \cref{sec:eva-trace}, the actual delay using \name is much lower.
}.
On the other hand, \name can also ensure the FCT degradation for new flows is bounded by an additive constant factor to the decay parameter ($\lambda$), which goes to negligible with the increase of the flow sizes.

\section{Occupancy-aware Flow Classification}
\label{sec:design-classify}

\begin{figure}
    \centering
    \begin{minipage}{.4\linewidth}
        \includegraphics[height=2.4cm]{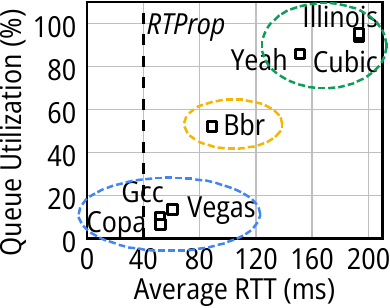}
        \caption{The relationship between queue utilization and delay in different CCAs. Experiments are simulated with real WiFi traces from~\cite{sigcomm2022zhuge}.}
        \label{fig:cca-avgrtt}
    \end{minipage}
    \begin{minipage}{.59\linewidth}
        \includegraphics[height=2.4cm]{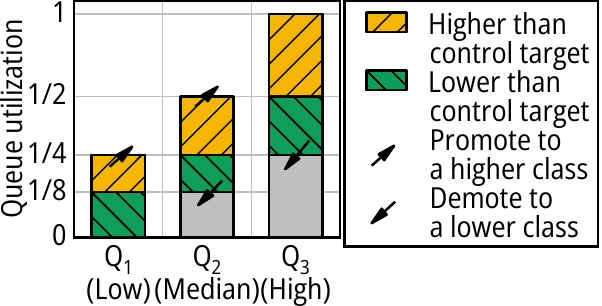}
        \caption{\name's hysteresis reclassification mechanism for flows. Only when the buffer occupancy of a flow has significantly deviated from the current class will it be moved to another class.}
        \label{fig:class-hysteresis}
    \end{minipage}
\end{figure}

As described in \cref{sec:design-overview}, \name seeks to classify flows into groups, each with a dedicated queue based on how aggressively they consume buffer space.
\textit{We find that flows implicitly demonstrate their preferences and objectives based on how they utilize the bottleneck queue}. 
We measure the buffer occupancy of 7 CCAs (the top-5 CCAs used in websites~\cite{sigmetrics2020gordon} plus two recent latency-sensitive CCAs, GCC and Copa), over real-world bandwidth traces (\cref{sec:eva-setup}).
We further measure the network RTT at the sender and the queue utilization on the bottleneck router.
A lower RTT indicates that this CCA is more latency-sensitive.
As we can see in \cref{fig:cca-avgrtt}, GCC, Copa, and Vegas have a low network RTT. 
Such CCAs achieve low latency by trying to keep the bottleneck queue as short as they can.
Real-time applications can choose these CCAs to achieve low latency.
In contrast, throughput-oriented CCAs (Cubic, Yeah, and Illinois) will maximize the queue utilization for high throughput.
This allows us to identify the latency sensitivity of flows by their queue occupancy: if one flow has a low queue occupancy at the bottleneck, it indicates that (i)~that flow tries to not overutilize the queue; and (ii)~that flow can co-exist with other flows with similar behaviors.

In this section, we present our hysteresis-based mechanism to robustly identify the flows~(\cref{sec:classify-hysteresis}) and our implementation considerations~(\cref{sec:classify-metric}).

\eat{

Having explained how \name gradually re-allocates bandwidth between old and new flows, we discuss how \name identifies the flows of interest.

Following the requirement \textbf{R3} in \cref{sec:motiv-req}, identifying flows and putting them into different queues without the labels from end hosts is essential and challenging. 
\name splits flows into queues according to their objectives on the premise that flows of similar performance objectives will not hurt each other.
To identify the objective of flows in the system, \name uses the queue occupancy. 

Having categorized flows according to their objective, the next natural question is \emph{(i)} how to split bandwidth across those categories; and \emph{(ii)} how long to wait before changing the bandwidth allocation.
For the former, our insight is that bandwidth allocation needs to depend on the ratio between the number of old and new flows. 
For the latter, our insight is to move bandwidth to new flows from old ones so fast as the old flows' CCA has time to react.
In practice, respecting the reaction time of each CCA means that we need to adapt the design of \name in various CCAs. 
To this end, we formulate the reaction procedure of different CCAs in \cref{sec:weight-theory} and figure out the reaction time of different latency-sensitive CCAs.
We could, therefore, design a uniform weight-adjustment algorithm for flows with different CCAs.
\name effectively focuses on the least reactive CCA to ensure all CCAs have adequate time to react.
}

\subsection{Hysteresis-based Adjustment}
\label{sec:classify-hysteresis}

\name puts short flows into a separate queue $Q_{new}$ and classifies long flows with different buffer occupancy aggressiveness into $Q_1,Q_2,\cdots,Q_n$.
Queue indices increase with buffer target \ie $Q_1$ will be shorter than $Q_3$, as shown in \cref{fig:overview}.
Each queue $Q_i$ targets a buffer occupancy of $q_0^{(i)}$.
We robustly classify flows as follows:

\parahead{Classification of new flows.}
The buffer aggressiveness of flow may take a long time to manifest.
Thus, \name will not characterize short flows lasting only a few RTTs~(\cref{sec:2}).
When the new flow is ready to be moved out from the new-flow queue $Q_{new}$ to one of the old queues (its weight reaching one, which we elaborated on in \cref{sec:weight-mechanism}), we measure the buffer occupancy of that flow $q_f$ \ie the number of packets of this queue that belong to flow $f$.
We then find the queue $i$ with the nearest target $q_0^{(i)}$ to accommodate this flow.

\parahead{Periodic adaptation.}
\name periodically examines flows and queues and moves flows accordingly in two steps.
While seemingly complex, these operations are well within the capabilities of Linux-based routers (\cref{sec:eva-impl}).

\ithead{Intra-queue examination} identifies outstanding flows among other flows in the current queue.
\name examines the buffer each flow occupies ($\frac{q_f}{\sum_{g\in Q_i}q_g}$) and its fair share ($\frac{1}{|Q_i|}$). 
If the buffer occupancy of a flow is larger than its fair share:
\begin{equation}
    \small
    \label{eq:share-increase}
    \frac{q_f}{\sum_{g\in Q_i}q_g} \geqslant \frac{1}{|Q_i|} + \alpha
\end{equation}
the flow is too aggressive in the current queue, where $\alpha > 0$ is a hysteresis.
\name wll promote that flow from queue $Q_i$ to $Q_{i+1}$ to keep $Q_i$ near its control target.
Similarly, a flow with an outstandingly lower buffer occupancy, i.e.:
\begin{equation}
    \small
    \label{eq:share-decrease}
    \frac{q_f}{\sum_{f\in Q_i}q_f} \leqslant \frac{1}{|Q_i|} - \alpha
\end{equation}
will be demoted from queue $Q_i$ to $Q_{i-1}$.
Here we set $\alpha$ to 10\% based on our previous observations in \cref{fig:cca-avgrtt}. 
Our evaluation in \S\ref{sec:eva} shows that the performance of \name is not sensitive to the workloads and CCAs.

\ithead{Queue-level examination} checks if the length of a queue fits the queue's control target.
 If the length of a queue exceeds a safe region between the control target of the neighbor queue, \name moves all flows in the current queue to that queue, as shown in \cref{fig:class-hysteresis}. 
 This is needed because the intra-queue examination only focuses on cross-flow \textit{relative occupancy}. 
 Thus, it cannot identify when flows in the current queue are comparably aggressive but more aggressive than the target of this queue.
For example, assume that two Cubic flows were previously classified to $Q_1$ (the least aggressive) due to being throttled elsewhere.
When these Cubic flows start to be aggressive, \name needs to move them to a different queue to protect incoming latency-sensitive flows.

\subsection{Design Considerations}
\label{sec:classify-metric}


In practice, \name has two following considerations.

\parahead{Number of queues to set.}
We observe that the CCAs are concentrated in three clusters (circles in \cref{fig:cca-avgrtt}).
Concretely, GCC, Copa, and Vegas have a queue occupancy of less than 20\%; Cubic, Illinois, and Yeah have a queue occupancy of more than 80\%; and BBR's stays in-between.
Therefore, we set three queues and use the average queue occupancy in these three clusters as our targets $\{q_0^{(i)}\}$.
We expect other CCAs to fall into one of these three representative categories, if not we can configure \name to work with more queues.


\parahead{Variation of buffer aggressiveness.}
A flow's buffer aggressiveness can change over time.
For example, a Cubic flow throttled/congested elsewhere (on a different router) will not be aggressive in buffer occupancy (although Cubic, the algorithm, would).
Such a Cubic flow can share the queue with other delay-sensitive flows.
However, when the bottleneck moves to the current router, this Cubic flow will be aggressive on the buffer occupancy, where the flow can no longer share the queue anymore.
Our reclassification mechanism is capable of correctly moving the flows, as evaluated in \cref{sec:eva-benchmark}.

\section{\name implementation}
\label{sec:eva-impl}

Implementing \name in Linux kernel has some challenges.
We discuss them and our solutions below.

\parahead{Order-preserving during reclassification.}
Flows can be moved to another class in the runtime.
Thus, we need to ensure the order-preservation during the reclassification of \name of a certain flow.
In response, we adopt a virtual class design in \name. 
During the enqueue process of new packets, we bind the \texttt{sk\_buff} to each flow.
During the dequeue process, we search for all flows that are bound to the determined class and dequeue the packet with the earliest enqueue time.
In this way, when moving a flow to another class, we can just rebind the pointer of the flow from the previous class to the new class.

\parahead{Reducing computational overhead.}
To implement \name in Linux kernel and optimize the execution overhead, we need to strictly optimize the computational overhead.
Specifically, we have the following two implementations:

\ithead{(i) Bit-shifts for exponential operations.}
\name reweights flows based on their ages with an exponential function, yet the floating number calculation in the kernel is expensive.
Therefore, we quantize the weight of new flows with the unit of $\frac{1}{128}$ and use bit shifts for the exponential weight updates, i.e., left shifting the weight by one bit every $\frac{1}{\lambda}$ milliseconds.

\ithead{(ii) Periodical reweighting and reclassification.}
The reweighting and reclassification are not necessary for each packet.
For the reweighting, we only need to reweight for a flow every $\frac{1}{\lambda}$ milliseconds.
When we set $\lambda=0.004$, this means to reweight every 250 ms.
For the reclassification, we should observe the results after moving one flow to a new class for at least one RTT to measure the queue utilization and observe the behavior of the flow in the new class.
Therefore, we also reclassify the flows in a periodic way -- we set the reclassification interval to 100ms.

\section{Evaluation}
\label{sec:eva}

\noindent We first present our experimental setup (\cref{sec:eva-setup}); then we evaluate \name by answering the following questions:
\begin{itemize}
    \item How does \name behave compared to baselines on real-world Web and bandwidth traces? \name reduce the stall duration of a real-time flow by 21\% to 87\% with various CCAs while maintaining comparable FCTs (\cref{sec:eva-trace}). 
    \item How sensitive is \name to workloads? \name is consistently performant with different sizes and numbers of Web flows, following our theoretical analysis (\cref{sec:eva-workload}).
    \item How does \name scale to multiple flows with different CCAs? We demonstrate that \name can correctly separate coexisting flows with different CCAs and provide consistent performance to all of them (\cref{sec:eva-classify}).
    \item How does \name perform in the testbed prototype? We implement \name in Linux kernel and show that \name reduces the stall duration by more than 60\% with reasonable overhead over real Web traces (\cref{sec:eva-testbed}).
    \item We further show that \name can outperform baselines when working with multiple real-time flows, bandwidth-probing CCAs, and different bottlenecks (\cref{sec:eva-benchmark}).
\end{itemize}

\begin{figure*}
    \centering
    \subfigure[When the RT flow uses Copa.]{
        \includegraphics[height=3.1cm]{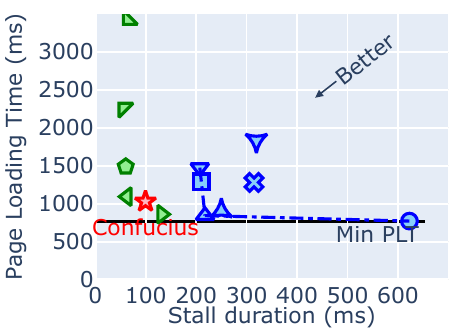}
        \label{fig:eva-tradeoff-copa}
    }
    \subfigure[When the RT flow uses GCC.]{
        \includegraphics[height=3.1cm]{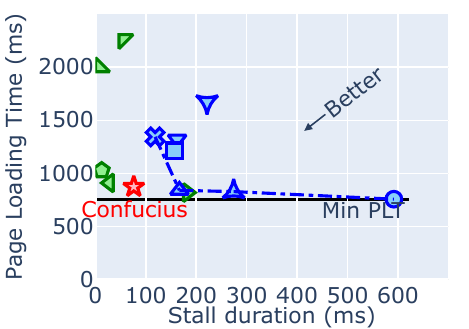}
        \label{fig:eva-tradeoff-gcc}
    }
    \subfigure[When the RT flow uses BBR.]{
        \includegraphics[height=3.1cm]{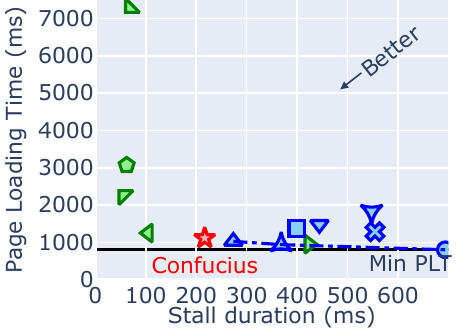}
        \label{fig:eva-tradeoff-bbr}
    }
    \subfigure{\includegraphics[height=3cm]{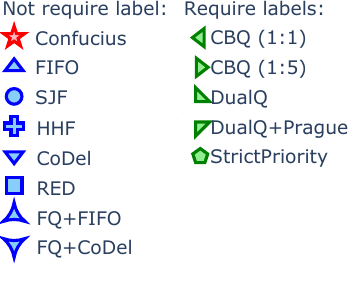}}
    \caption{The trade-off between the real-time (RT) flow (stall duration) and Web flows (page loading time) on bandwidth traces \texttt{C1}. The dashed line denotes the Pareto front of baselines that do not require labels from end hosts. We mark baselines in green if they rely on labels from end hosts, and in blue if not. We change the CCA that the real-time flow uses in different subfigures.}
    \label{fig:eva-tradeoff-real}
\end{figure*}

\subsection{Experiment Setup}
\label{sec:eva-setup}

\parahead{\texttt{Ns-3} setup.}
In \cref{sec:eva-trace,sec:eva-workload,sec:eva-classify}, we evaluate the performance of \name with \texttt{ns-3.34}.
We use the example in \cref{fig:intro-example} and limit the capacity of the bottleneck link based on the bandwidth traces from~\cite{sigcomm2022zhuge}.
The dataset contains 3 sets of cellular traces (\texttt{C1}: mixed; \texttt{C2}: 4G; \texttt{C3}: 5G), and 2 sets of WiFi traces (\texttt{W1}: Office; \texttt{W2}: Restaurant), where the average bandwidth ranging from 22 Mbps to 375 Mbps (details in \cref{app:eva-traces}).
The round-trip propagation delay is set to 40ms based on~\cite{sigcomm2022zhuge}.
We adopt the RTC library in \texttt{ns-3} from~\cite{nsdi2024hairpin, sigcomm2022zhuge}.
We evaluate the real-time flow with different delay-sensitive CCAs, including Copa~\cite{nsdi2018copa} (used by Meta Live~\cite{flive-quic}), GCC~\cite{ton2017webrtc} (used by WebRTC), and BBR~\cite{queue2016bbr}. 
The Web flows use the mostly deployed CCA~\cite{sigmetrics2020gordon} -- Cubic~\cite{sigops2008cubic}.

\parahead{Web traces.}
To compose a realistic and relevant dataset of Web traffic, we collected the Alexa Top-1000 websites\footnote{
    Although the Alexa Top website list has been deprecated, we still use this list since it is the most well-known list for top websites. 
}~\cite{alexa-top1k}.
We use \texttt{selenium}~\cite{selenium} to automatically load Web pages through Google Chrome (version 120.0.6099.218), and use \texttt{NetLog}~\cite{netlog} and \texttt{browsermobproxy}~\cite{browsermob} to record the packets and socket states.
The measurement was run in November 2023, with distribution in \cref{fig:web-measure2}. 
The version of HTTP is negotiated with the website, where the majority is \texttt{HTTP/1.1} and \texttt{HTTP/3.0}.
We show in Appx.~\ref{app:web} that although the parallel connection of \texttt{HTTP/1.1} contributes to the concurrency, the majority of concurrency still comes from the diverse objects on the Web page.
We replay the Web traces to test a variety of scenarios.

\parahead{Baselines.}
We compare \name with multiple schedulers, categorized and listed below. 
We use the default parameters in the Linux kernel 4.4.0 and \texttt{ns-3.34} for baselines.

\ithead{Not require labels (Note that \name does not require either):}
\begin{itemize}[leftmargin=6mm]
    \item[(1)] \texttt{FIFO} and (2) \texttt{FQ+CoDel}~\cite{rfc8290fqcodel}, the default \textsf{qdisc}s in Linux (before and after \textsf{systemd} v217~\cite{systemd-qdisc}). 
    \item[(3)] \texttt{FQ+FIFO} is the fair queueing without the AQM. 
    \item[(4)] \texttt{CoDel}~\cite{cacm2012codel} and (5) \texttt{RED}~\cite{ton1993red} will drop packets before the queue overflows to notify the sender.
    \item[(6)] \texttt{SJF} (shortest job first) prioritizes short flows over long flows, which is exactly opposite to what \name tries to do. We take the implementation from PIAS~\cite{nsdi2015pias}.
    \item[(7)] \texttt{HHF}~\cite{sigcomm2002hhf} (heavy-hitter filter) differentiates between small flows and heavy-hitters and schedules separately.
\end{itemize}
\ithead{Require labels:}
\begin{itemize}[leftmargin=6mm]
    \item[(8)] \texttt{DualQ}~\cite{rfc9332dualq} is a recently proposed scheduler in L4S~\cite{rfc9330l4s} that protects latency-sensitive flows with labels, using the DSCP bits to identify the traffic and notify the sender.
    \item[(9)] \texttt{DualQ+Prague}. \texttt{DualQ} provides ECN signals and works best with TCP Prague~\cite{briscoe-iccrg-prague-congestion-control-03} in the L4S framework by design. We adopt the implementation from~\cite{prague-ns3}.
    \item[(10)] \texttt{CBQ (1:1)} and (11) \texttt{CBQ (1:5)} are the weighted class-based queues, which put flows into different classes based on application labels. We set the weights for two classes (RT:Web) to 1:1 and 1:5 and evaluate respectively.
    \item[(12)] \texttt{StrictPriority} strictly prioritizes traffic from real-time flows if they are labeled accordingly.
\end{itemize}

\parahead{Metrics.}
We focus on the following metrics in experiments.
\begin{itemize}
    \item \textit{Stall duration} for video frames is the duration for which the delay of the video frame is greater than 190 ms. This reflects users' experiences on video stalls~\cite{itu-ddl, sigcomm2022zhuge, infocom2022dams}. We use this metric to evaluate how the RT flow is affected.
    \item \textit{Page Load Time (PLT)} is the time till the last HTTP request in a web page is completed. We use this metric to evaluate the performance of web traffic. PLT degradation refers to the increase of delay compared to \texttt{FQ}.
\end{itemize}
Besides, we also evaluate other metrics in different experiments, which we will elaborate on accordingly.

\subsection{\name under a realistic workload}
\label{sec:eva-trace}

\parahead{Simulation scenario.}
We have a long-running real-time flow from the RTC module in ns-3.
We then randomly select a website and replay the traces we collected to compete with the real-time flow.
We set the interval of loading two websites to 53 seconds, which is the average Web page viewing time~\cite{page-stay}.
In each run, we measure the duration where the frame delay of the video flow is larger than 190 ms (stall).
We also measure the Web PLT for websites.
We repeat the experiment for three CCAs for the real-time flow.
In this subsection, we present the results over \texttt{C1} bandwidth traces and leave others to Appx.~\ref{app:eva-traces}.
We present the average PLTs and stall durations in \cref{fig:eva-tradeoff-real}, and dive into distributions later.

\parahead{\name strikes a balance between video and web performance that is consistent across CCAs.}
In \cref{fig:eva-tradeoff-copa}, we observe that schedulers not relying on labels from the end host (marked in blue) suffer from long video stalls. 
For example, when the real-time flow adopts Copa, using \texttt{FQ-CoDel} or \texttt{FIFO}, the real-time flow experiences a stall of 200 ms to 300 ms on average when loading different websites.
AQMs such as CoDel will further impair the performance of the Web flows since AQMs will drop packets for those flows.
In contrast, \name can reduce the stall duration to less than 100 ms and improve by more than a half against all other baselines not requiring baselines. 

Schedulers requiring labels (marked in green) protect prelabeled RT flows, but considerably degrade the PLT for the Web traffic. 
\texttt{DualQ+Prague} improves \texttt{DualQ} since the CCA on the end-host can react more effectively, but still incur considerable penalty on Web flows since it incurs packet drops to the Web flows (non-scalable from DualQ's design).
Note that even when using \texttt{StrictPriority}, the real-time flow still suffers from 60 ms degradation on average due to bandwidth fluctuation.
\name is almost on par with schemes relying on end-host labels in terms of the delay of the real-time flows.
Remember that it is unrealistic to assume that an end-host will correctly label all traffic (\cref{sec:motiv-related}).

Most importantly, \name does not incur too much penalty on the PLT of Web pages, and pushes the Pareto front of the schedulers not requiring labels (the dashed blue line) forward.
\name reduces the PLT compared to \texttt{CoDel}, \texttt{RED}, \texttt{FQ+CoDel}, and \texttt{HHF} since \name gently adjusts the bandwidth share for the Web flows.
Even compared to FQ+FIFO, \name only increases the average PLT by \textit{up to 8\% (up to 88 ms)} in three subfigures, which is much smaller than the improvement on the stall duration.
The improvement when using BBR is not as significant as the other two CCAs -- this is due to the suboptimal performance of BBR in controlling the delay for the real-time flow, as we also saw in \cref{fig:cca-avgrtt}.

\begin{figure}
    \centering
    \subfigure[Number of websites that lead to non-zero stalls for the RT flow.]{
        \includegraphics[height=3.2cm]{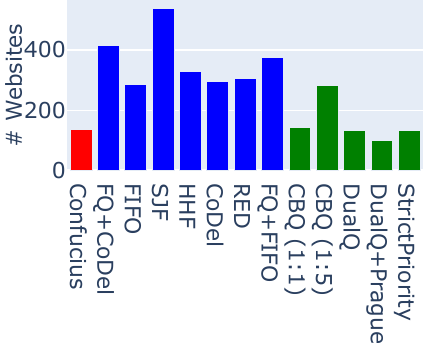}
        \label{fig:num-delay-dur}
    }\hfill
    \subfigure[Number of websites that suffer from PLT of longer than 2 seconds.]{
        \includegraphics[height=3.2cm]{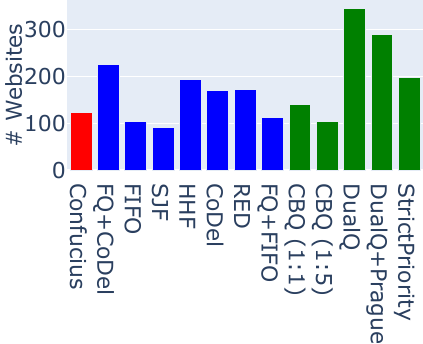}
        \label{fig:num-plt}
    }
    \caption{The number of runs, out of 1000 websites, that lead to the degradation of the real-time and Web flows. The lower the better. \cref{fig:num-delay-dur} cuts \cref{fig:cdf-copa-delay} at 190 ms.}
    \label{fig:eva-num-copa}
\end{figure}

\parahead{\name protects the real-time flow when competing with traffic from 86\% of the websites.}
We further break down the details for different websites in \cref{fig:eva-tradeoff-copa} into \cref{fig:eva-num-copa}.
In \cref{fig:num-delay-dur}, we present the number of websites that do not affect the delay of the real-time flow (the stall is negligible of shorter than 190 ms).
The lower the number is, the better performance the scheduler is.
When using \name, the real-time flow will only suffer from stall when competing with 136 out of 1000 websites.
However, for all other baselines that do not require labels, the number ranges from 288 websites (\texttt{FIFO}) to 537 websites (\texttt{SJF}).
\name reduces the number by 53\%-75\%.
Even for baselines requiring labels excluding \texttt{CBQ (1:5)}, the number ranges from 101 to 143.
This further shows that \name can achieve comparable or sometimes even better protection to the real-time flow as those label-based solutions.

We also measures the number of websites suffering from a long PLT of longer than 2 seconds, which is the threshold for good user experience~\cite{sigcomm2019e2e}.
When using \texttt{FQ+CoDel}, 227 websites will suffer from long PLT, while \name can reduce this number by half to 127.
\name's results is comparable to \texttt{FQ+FIFO}, demonstrating the fairness of \name for competing Web flows.
For baselines requiring labels that behave well in \cref{fig:num-delay-dur}, at least 198 websites suffer from long PLTs.
The closest label-based baseline is \texttt{CBQ (1:1)}, which we can also see from \cref{fig:eva-tradeoff-real}. 
Besides the unrealistic label requirement, we will later show in \cref{sec:eva-workload} that \texttt{CBQ (1:1)} does not scale to the variation of workloads.
\cref{fig:eva-tradeoff-real} evaluates with Web flows, but if the competing flows are not Web flows but FTP flows instead, the performance will be drastic since \texttt{CBQ (1:1)} adopts a fixed ratio between classes.

\begin{figure}
    \centering
    \subfigure[Stall duration of the RT flow.]{
        \includegraphics[height=2.7cm]{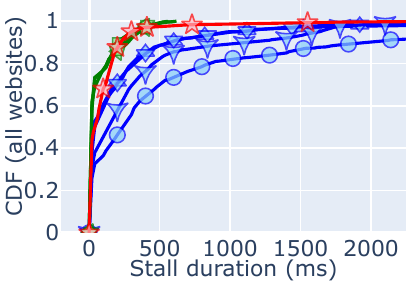}
        \label{fig:cdf-copa-delay}
    }\hfill
    \subfigure[PLT degradation of Web flows, compared to FQ.]{
        \includegraphics[height=2.7cm]{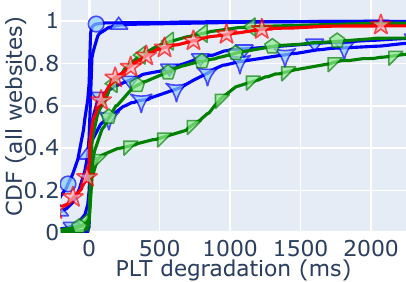}
        \label{fig:cdf-copa-plt}
    }\hfill
    \subfigure[The max frame delay of the RT flow when Web flows arrive.]{
        \includegraphics[height=2.7cm]{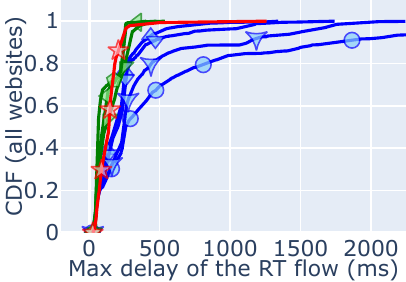}
        \label{fig:cdf-copa-maxdelay}
    }\hfill
    \subfigure[The delay distribution of all RT packets during the competition.]{
        \includegraphics[height=2.7cm]{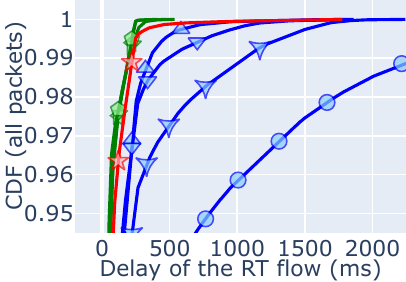}
        \label{fig:cdf-copa-alldelay}
    }
    \caption{The distributions in \cref{fig:eva-tradeoff-copa}. The legend is the same as \cref{fig:eva-tradeoff-real}. We present a part of the baselines for simplicity.}
    \label{fig:eva-cdf-copa}
\end{figure}

\begin{figure*}
    \centering
    \subfigure[The frame delay of different flows.]{
        \includegraphics[height=3.2cm]{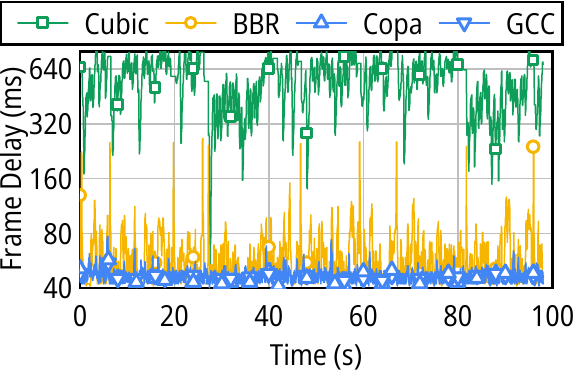}
        \label{fig:delay-timeline}
    }
    \subfigure[The classification results in different time.]{
        \includegraphics[height=3.2cm]{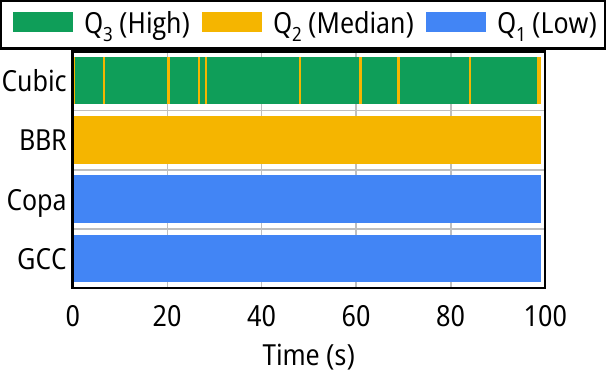}
        \label{fig:classify-results}
    }
    \subfigure[The JFIs and delays among baselines.]{
        \includegraphics[height=3cm]{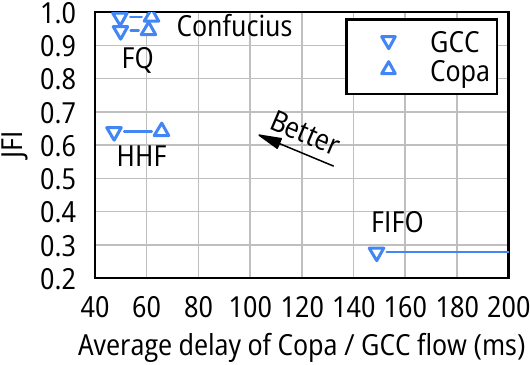}
        \label{fig:jfi-mindelay}
    }
    \caption{Four flows with different CCAs (Cubic, BBR, Copa, and GCC) run in the same bottleneck router. We present the frame delay and classification results of these flows when using \name over time in \cref{fig:delay-timeline} and \ref{fig:classify-results}. We also compare the fairness (JFI) and the delay of latency-sensitive flows (Copa and GCC) of \name and baselines in \cref{fig:jfi-mindelay}.}
    \label{fig:four-flow}
\end{figure*}

\parahead{\name controls the delay and PLT following the theoretical analysis.}
\cref{fig:cdf-copa-delay} further presents the distribution of stall duration when the video flow encounters Web flows from different websites in the dataset.
With \texttt{FQ+CoDel} or \texttt{FIFO}, the stall for the real-time flow will last for longer than 500 ms for 12\% (\texttt{FIFO})-18\% (\texttt{FQ+CoDel}) websites, where the number for \name is 1\%.
In contrast, with \name, the real-time flow will not experience any stall when encountered with 95\% of the websites, comparable to \texttt{CBQ}.
Importantly, besides the PLT measured in \cref{fig:num-plt}, \name does not over-penalize web traffic -- 60\% of websites will not suffer from a penalty at all against \texttt{FQ}, as shown in \cref{fig:cdf-copa-plt}, which mostly corroborates our previous theoretical analysis.
We further present the distribution of maximum experienced delay for the real-time flow in \cref{fig:cdf-copa-maxdelay}.
The fraction of having a maximum delay of $>$500 ms is 1\% using \name, while for \texttt{FIFO} and \texttt{FQ+CoDel} are 5\% and 18\%.
This further demonstrates that \name can control the latency fluctuation in not only the stall duration but also directly the raw delay.
The dive into the network delay of all packets of the real-time flow in \cref{fig:cdf-copa-alldelay} corroborates this as well.
The results when using GCC and BBR are similar.

\subsection{\name under workload changes}
\label{sec:eva-workload}

\begin{figure}
    \centering
    \subfigure[Stall duration of the existing \newline real-time flow.]{
        \includegraphics[height=2.45cm]{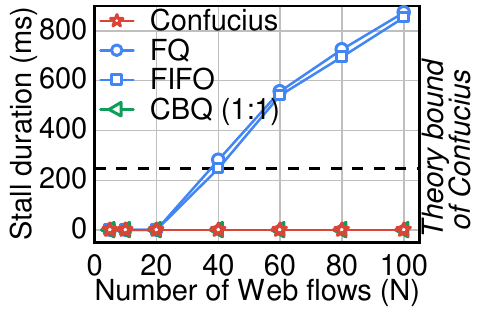}
        \label{fig:shortn-delay}
    }
    \subfigure[PLT degradation of Web (new) flows against \texttt{FQ}.]{
        \includegraphics[height=2.45cm]{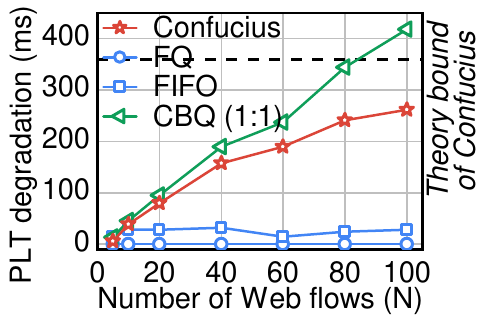}
        \label{fig:shortn-fct}
    }
    \caption{Performance consistency in workloads with different \textit{number} of Web flows, each flow with the size of 15KB.}
    \label{fig:eva-shortn}
\end{figure}

\begin{figure}
    \centering
    \subfigure[Stall duration of the existing \newline real-time flow.]{
        \includegraphics[height=2.39cm]{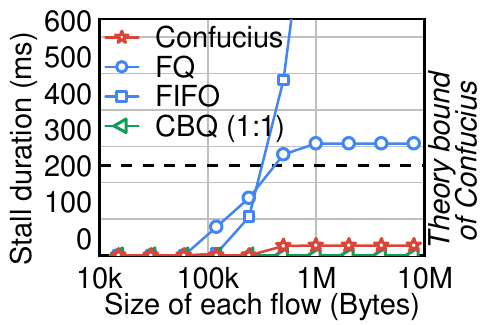}
        \label{fig:bsize-delay}
    }
    \subfigure[PLT degradation of Web (new) flows against \texttt{FQ}.]{
        \includegraphics[height=2.39cm]{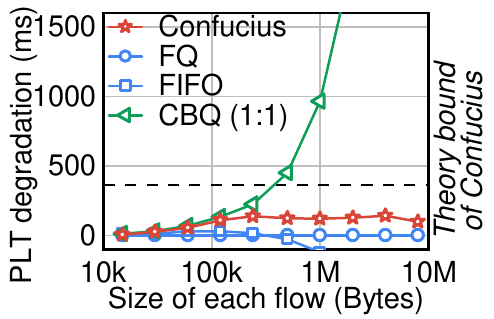}
        \label{fig:bsize-fct}
    }
    \caption{Performance consistency in workloads with different \textit{size} of Web flows, each experiment having 5 flows. The dashed line is the theoretical bounds from \cref{tab:theory-summary}.}
    \label{fig:eva-bsize}
\end{figure}

In this subsection, we test our theoretical analysis by investigating whether \name can provide consistent performance in controlled workloads. 
We vary the workload by changing the number of flows in a Web page and the size of Web flows. 
We measure the stall duration in different scenarios and the degradation on the PLT against \texttt{FQ}. 

\parahead{\name is bounded by theoretical thresholds}, confirming our analysis.
We vary the number of Web flows from 5 to 100, each with the size of 15KB (medium flow size in our measurement), and summarize our results in \cref{fig:shortn-delay}. 
The stall duration for \texttt{FQ} and \texttt{FIFO} increases with the number of flows: when the number of Web flows goes to 60, the real-time flow experiences a stall for more than half a second.
On the contrary, \name maintains zero stall in this setting, similar to \texttt{CBQ}.
We further compare the experimental results with our previous analysis in \cref{sec:weight-theory}.
As we can see in the dashed line in \cref{fig:eva-shortn}, the experimental results corroborate our theoretical analysis for \name in \cref{tab:theory-summary}.

We further change the size of Web flows (from short flows to long flows) and see if \name is capable of handling all types of competing traffic.
With the increase of the size of flows, the competing flows are changing from short flows (e.g., Web) to long flows (e.g., FTP).
We vary the size of Web flows from 15KB to 9MB, and run 5 flows with the same size to compete with the HRT flow. 
When using \texttt{FIFO}, the real-time flow will suffer from drastic stall due to failure to provide inter-CCA fairness across flows, as shown in \cref{fig:bsize-delay}.
The \elephant flow using \texttt{FQ} also has a long stall of hundreds of milliseconds.
In contrast, \name is still able to achieve negligible stall for the \elephant flow and bounded PLT degradation for Web flows at the same time.

\subsection{Heterogeneous Flow Classification}
\label{sec:eva-classify}

In this subsection, we zoom in on \name's flow classification mechanism. 
We find that \name can accurately group flows of the same/similar CCA together without any prior knowledge or labels from end hosts, which in turn leads to better performance compared to the baselines.

We simultaneously run four long flows of different CCAs: one Cubic flow, one BBR flow, one GCC flow, and one Copa flow for 100 seconds.
We plot the network delay for each flow over time in \cref{fig:delay-timeline}.
We can clearly see that Copa and GCC enjoy a consistent low latency around 40-60 ms, even when they are competing with BBR and Cubic flows.

To understand \name's superior performance, we look at its classification results over time and present them in \cref{fig:classify-results}.
Four bars represent the classification results of \name for four flows over time, while three colors indicate which queue the flow is classified into.
\name can correctly classify flows using different CCAs into the correct queues: Copa and GCC flows can be stably put into the low occupancy queue ($Q_1$, blue), the BBR flow into the median occupancy queue ($Q_2$, yellow), and the Cubic flow into the high occupancy queue ($Q_3$, green).
This follows our previous observation in \cref{fig:cca-avgrtt} -- Copa and GCC both demonstrate similar low buffer occupancy, while Cubic occupies the buffer aggressively, and BBR in the middle.
Moreover, we notice that the Cubic flow can temporarily be in the same queue as BBR, as shown in the yellow lines in the green bar in \cref{fig:classify-results}.
This is expected as the Cubic flow has (at times) a low queue occupancy in its probing period.
Second, flows with different CCAs can co-exist in the same queue as long as they have similar buffer occupancy.
In this experiment, Copa and GCC flows are put into the same queue since they have similar buffer occupancy.
As we can see in \cref{fig:delay-timeline}, these two flows still have consistent low latency all the time.

We also measure the Jain's fairness index (JFI) in \cref{fig:jfi-mindelay} to present the fairness when using different schemes. 
We compare the results (the delay of the Copa and GCC flow, and the JFI of all flows) in the same experiment with other schedulers in \cref{fig:jfi-mindelay}.
With \name, the Copa and GCC flows also enjoy a reasonable fair share of the bandwidth as \texttt{FQ} -- the JFI in this experiment is 0.98 in \cref{fig:jfi-mindelay}, where JFI close to one indicates a better fairness.

\subsection{Testbed Experiments}
\label{sec:eva-testbed}

\begin{figure}
    \centering
    \subfigure[The real-time flow's stall vs. Web flows' load time.]{
        \includegraphics[height=2.5cm]{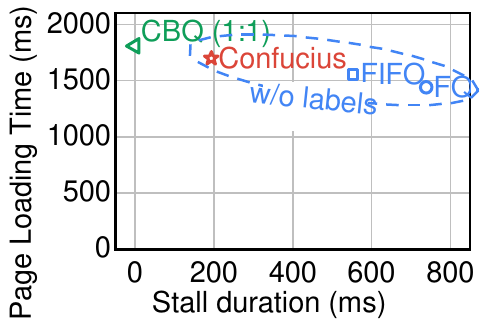}
        \label{fig:tradeoff-kernel-copa}
    }
    \subfigure[Processing time for each packet. Axes are log-scaled.]{
        \includegraphics[height=2.5cm]{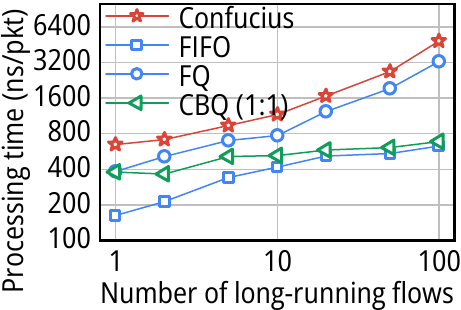}
        \label{fig:time-kernel}
    }
    \caption{Results over our Linux kernel-based testbed.}
    \label{fig:eva-kernel}
\end{figure}

We implement \name as a kernel module of queue disciplines (\texttt{qdisc}) in traffic control in Linux kernel 4.4.0 (1.4k LoCs) and evaluate the performance of \name on a software router based on Intel Xeon E5-2620 v4 CPU.
We run the official implementation of Copa~\cite{copa-ccp}.
We find that \name achieves significant benefits in kernel implementations while only adding marginal processing delay.

We stream the video frames using socket and TCP Copa, and measure the end-to-end delay for the real-time flow.
We then set up an HTTP server based on Python to replay the Web traces we collected.
We also measure the computational overhead of \name and the baselines.
We record the processing time for the enqueue and dequeue operation in Linux \texttt{tc} using \texttt{printk}, where the reweight and reclassification in \name are both implemented.

As shown in \cref{fig:tradeoff-kernel-copa}, \name reduces the stall duration by more than 60\% without the need for labels on each packet.
This result is similar to our simulation in \cref{fig:eva-tradeoff-copa}.
Moreover, from our experiments, 86\% of websites when using \name do not suffer from stall at all. 
In contrast, this number is only 56\% and 30\% for \texttt{FIFO} and \texttt{FQ}.

We vary the number of long-running flows to measure the overhead of \name.
Note that the processing time of \name is insensitive to the number of short flows, as they all belong to the new-flow queue.
As shown in \cref{fig:time-kernel}, \name slightly increases the processing time for each packet compared to \texttt{FQ}.
Even for 100 concurrent long-running flows, the per-packet processing time of \name is still 5 $\mu s$, indicating a processing rate of 200 kpps and is at the same magnitude as \texttt{FQ}.
Since \name is mainly designed for last-mile routers such as home routers, this can satisfy the daily usage of home access points or last-mile routers.
We stress that the kernel implementation of \name can be further optimized for high-performance execution in the future.
We leave the further exploration of \name over numerous flows (\eg in the routers in the core network) in the future.

\subsection{Microbenchmarks}
\label{sec:eva-benchmark}

We further evaluate the performance of \name in a series of microbenchmarking settings.
In Appx.~\ref{app:eva-probing}, we demonstrate that the hysteresis mechanism of \name (\S\ref{sec:classify-hysteresis}) can work with bandwidth-probing CCAs (e.g., BBR) and stably and correctly classify flows.
We further show that \name will not have side effects on the fairness aspect in Appx.~\ref{app:eva-jfi}, and investigate if the bottleneck is not the router where \name is deployed in Appx.~\ref{app:eva-btlbw}.
Even if multiple real-time flows are competing simultaneously, \name can still handle those flows and provide significant performance improvements against baselines (Appx.~\ref{app:eva-multivideo}).
\section{Limitations}
\label{sec:limit}

We discuss some other related work besides \cref{sec:motiv-cause} in Appx.~\ref{app:related}, and outline some limitations of \name here.

\parahead{Applications using latency-sensitive CCAs.}
In this paper, we assume that real-time applications will use latency-sensitive CCAs.
For example, video conferencing applications will use CCAs such as GCC and Copa but not Cubic~\cite{sigcomm2022zhuge}.
This general holds since the operators of applications will optimize towards their goal, where the latency is definitely the goal of real-time applications.
If the application does not follow this, that means the application's CCA itself is still problematic, and the effect of \name will be limited.

\parahead{Web flow characteristics for mobile applications.}
The measurement of Web flows in \cref{sec:2} and Appx.~\ref{app:web} is loading Web pages from desktop browsers (Google Chrome).
We do not conduct measurements over mobile Apps due to device limitation.
However, the root cause contributing to the bursts still exists in mobile Apps -- one page contains diverse objects (images, videos, scripts) from different domains.
We leave the investigation of Web pages on mobile Apps to future.

\parahead{\name scales to core routers.}
We mainly discuss the bottleneck at the edge routers in this paper.
This is because when the network delay increases, it is more likely to happen at the edge~\cite{sigcomm2022zhuge}.
In core routers, due to high line rate, 100 new flows is not a big number and will likely not result in drastic available bandwidth fluctiation. 
Meanwhile, tracking per-flow state such as \texttt{FQ} is already burdensome, therefore is out of the scope of \name.
Nevertheless, encouraged by recent sophisticated AQM on high-performance switches~\cite{sigcomm2022abm}, it might be possible to extend \name to core routers.


\section{Conclusion}
\label{sec:concl}

\noindent We propose \name, the first queue management scheme to protect the real-time flows in the flow competition while not requiring any labels from end hosts.
\name achieves this by grouping flows based on their latency preferences, which it infers by observing their buffer occupancy over time. \name gradually adjusts the service rates of flows to match the reaction of congestion control. 
Doing so allows \name to mitigate latency spikes of real-time flows.
Extensive evaluation shows that \name protects the real-time flows from stalls when competing with 86\% websites, almost doubling over numerous baselines.

This work does not raise any ethical issues.

\bibliographystyle{plain}
\bibliography{bibfile}

\appendix



\section{Web Page Connection Analysis}
\label{app:web}

\begin{table}
    \small
    \centering
    \begin{tabular}{ccccc}
    \hline
        Rank & Website & \texttt{ACTIVE} & \texttt{IN\_USE} & \texttt{OPEN} \\
    \hline
        189 & \rurl{dailymail.co.uk} & 50 & 134 & 250 \\
        109 & \rurl{tumblr.com} & 39 & 82 & 153 \\
        89 & \rurl{w3schools.com} & 32 & 95 & 176 \\
        147 & \rurl{speedtest.net} & 28 & 87 & 137 \\
        113 & \rurl{cnn.com} & 27 & 118 & 194 \\
        186 & \rurl{namu.wiki} & 27 & 112 & 192 \\
        173 & \rurl{indiatimes.com} & 22 & 99 & 136 \\
        106 & \rurl{rakuten.co.jp} & 20 & 68 & 97 \\
        35 & \rurl{fandom.com} & 19 & 68 & 97 \\
        7 & \rurl{yahoo.com} & 19 & 42 & 63 \\
    \hline
    \end{tabular}
    \caption{Websites in Top 200 that have the highest number of \texttt{ACTIVE} flows.}
    \label{tab:web-top-active}
\end{table}

\begin{figure}
    \centering
    \includegraphics[width=\linewidth]{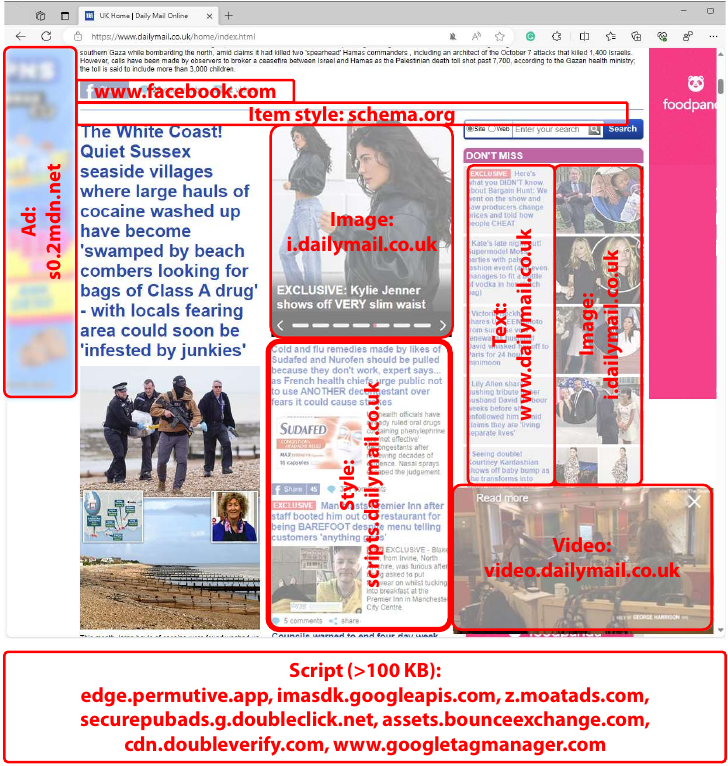}
    \caption{The screenshot of\rurl{dailymail.co.uk}, loaded in November 2023.}
    \label{fig:dailymail-home}
\end{figure}

\parahead{Many \texttt{ACTIVE} flows.}
In this section we provide more details about the Web traces we measured in \cref{sec:2}.
\cref{tab:web-top-active} presents the websites that have the highest number of concurrent \texttt{ACTIVE} flows.
We can see that\rurl{dailymail.co.uk} has 50 concurrent \texttt{ACTIVE} flows, and 250 \texttt{OPEN} sockets at the same time.
This is due to the complicated page structure of the homepage of\rurl{dailymail.co.uk}.
We present a screenshot of the homepage of\rurl{dailymail.co.uk} in \cref{fig:dailymail-home}.
We can clearly see that there are many objects on the homepage, visible (images, texts, videos) and invisible (scripts, styles).
Some objects have dependency over others, so the concurrent \texttt{ACTIVE} flows are fewer than concurrent \texttt{OPEN} sockets, but that still result in 50 flows.

\begin{figure}
    \centering
    \subfigure[Connections per Web page.]{
        \includegraphics[height=2.6cm]{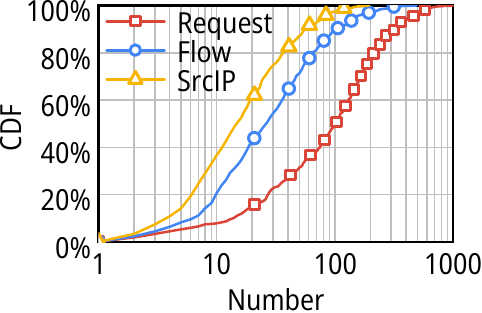}
        \label{fig:web-cnt}
    }
    \subfigure[Size distributions.]{
        \includegraphics[height=2.6cm]{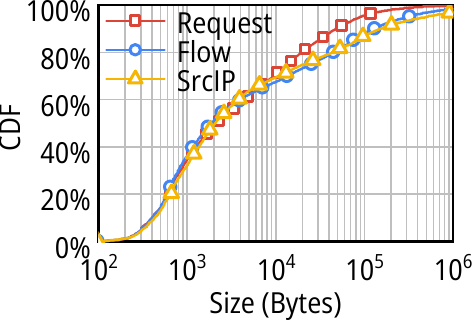}
        \label{fig:web-size}
    }
    \caption{Number of unique IPs/requests and their size for loading each of Alexa top 1000 websites.}
    \label{fig:web-measure}
\end{figure}

\parahead{Domains and requests.}
We further present the distribution of the number of unique HTTP requests and source IPs in \cref{fig:web-cnt}, together with the size distribution in \cref{fig:web-size}.
The median number of unique IPs that loading the homepage of a website will request is 15, while flow sizes range from 100 bytes to 100KBytes with a median number of 15KB.
This is also the size of Web flows we used in \cref{sec:eva-workload}.

\begin{table}
    \small
    \centering
    \begin{tabular}{ccccc}
    \hline
         & \texttt{HTTP/1.0} & \texttt{HTTP/1.1} & \texttt{HTTP/2.0} & \texttt{HTTP/3.0}\\
    \hline
        By website & \makecell{40\\(2.37\%)} & \makecell{961\\(56.97\%)} & \makecell{13\\(0.77\%)} & \makecell{673\\(39.89\%)}\\
    \hline
        By request & \makecell{49\\(0.02\%)} & \makecell{178341\\(87.29\%)} & \makecell{75\\(0.04\%)} & \makecell{25833\\(12.64\%)}\\
    \hline
    \end{tabular}
    \caption{The distribution of HTTP versions counted by websites and requests. The sum of websites is greater than 1000 since loading one websites may generate requests to different domains, which can use different HTTP versions.}
    \label{tab:web-http-ver}
\end{table}

\parahead{Composition of HTTP versions.}
A straightforward understanding of multiple flows of loading one Web page is the parallel connection introduced in HTTP/1.1.
Our measurement in \cref{fig:web-cnt} does show the effect of parallel connection -- the median number of connections counted by flow is 2x of that of source IPs.
However, the root cause is still the diverse objects on one page, as shown in \cref{fig:dailymail-home}.
To help to better understand the composition of HTTP requests, we present the distribution of HTTP versions in \cref{tab:web-http-ver}.
We can see that different websites actually have a very diverse structure of HTTP versions, where the majority is HTTP/1.1 and HTTP/3.0.

\section{Fluid Model Analysis}
\label{app:theory}

In this section, we present the details about how we get the results in \cref{tab:theory-summary}.
We list the notations we will use in \cref{tab:theory-notation}.

\begin{table}
    \centering
    \small
    \begin{tabular}{cl}
        \hline
        \multicolumn{2}{l}{\textbf{Parameters and variables:}} \\
        $B$ & Size of each new Web flow.\\
        $N$ & Number of new Web flows.\\
        $k$ & The responsiveness of a CCA.\\
        $q_0$ & The delay target that a CCA will try to achieve.\\
        $C$ & The link capacity.\\
        $\tau$ & The feedback loop of a CCA (usually one RTT).\\
        $B_0$ & The initial burst of a new flow (\eg the initial \texttt{cwnd}~\cite{ccr2010initcwnd}).\\
        $P$ & The scheduling policy.\\
        \hline
        \multicolumn{2}{l}{\textbf{Functions:}} \\
        $s(t)$ & Sending rate of the \elephant flow of time $t$.\\
        $r(t)$ & Available bandwidth of the \elephant flow of time $t$.\\
        $p(t)$ & Number of packets in the queue of the \elephant flow.\\
        $q(t)$ & The queueing delay of the \elephant flow.\\
        \hline
    \end{tabular}
    \caption{Notations}
    \label{tab:theory-notation}
\end{table}

\parahead{CCA model.}
We adopt a simplified delay-convergent CCA model~\cite{sigcomm2022starvation, sigmetrics2011stability}, where the delay-sensitive CCA has a target queueing delay, $q_0$.
The CCA seeks to maintain its queueing delay around this target, increasing or decreasing its sending rate proportional to the difference between the current delay and the target:
\begin{equation}
    \small
    \label{eq:model}
    \frac{{\rm d}s(t)}{{\rm d}t} = -k\cdot (q(t-\tau)-q_0)
\end{equation}
Here, $s(t)$ and $q(t)$ are the flow's instantaneous sending rate and queueing delay at time $t$, and $\tau$ is the feedback loop of the CCA.
$k$ is a coefficient representing the CCA's responsiveness, indicating how aggressive the CCA is when the delay changes.
We explain it quantitatively in Appx.~\ref{app:theory-responsive}. 

\parahead{Delay model.}
Next, we analyze the number of packets in the queue, $p(t)$, at time $t$.
At any $t>0$, this quantity satisfies the following relationship:
\begin{equation}
    \small
    \label{eq:buffer}
    p(t) = p(0) + \int_0^t\left(s(t')-r(t')\right){\rm d} t'
\end{equation}
where $p(0) = q_0\cdot C$ is the number of packets in the buffer in steady state with $C$ being the link capacity.
If $r(t)$ represents the service rate for the \elephant flow at time $t$, then the queueing delay can be written as follows:
\begin{equation}
    \small
    \label{eq:delay}
    q(t) = \frac{p(t)}{r(t)} 
\end{equation}

The real-time flow and the competing flows focus on different metrics.
The real-time flow focuses on the \textbf{maximum queueing delay}, $q^{max}_{P}$, for a given scheduling policy $P$:
\begin{equation}
    \small
    \label{delay-all}
    q^{max}_{P} = \max_{t>0}\ q(t)
\end{equation}
In this context, we find that $q^{max}_{P}$ serves as a good proxy for the duration of delay degradation since it establishes a \textit{lower bound} on how quickly previously-queued packets of the \elephant flow drain from the bottleneck queue.

The Web flows will focus on flow completion time (\textbf{FCT}), $T$, which can be expressed as follows:
\begin{equation}
    \small
    \label{eq:fct-all}
    \int_0^T\left(C-r(t')\right){\rm d}t'=N\cdot B
\end{equation}

Having established our two figures of merit (\textit{maximum queueuing delay} and \textit{FCT degradation to FQ}), we evaluate four scheduling policies: FQ, FIFO, CBQ (1:1), and \name. 
We find that the available bandwidths for these policies satisfy the following relationships:
\begin{subequations}
\small
\label{eq:rrate}
\begin{align}
    r_{FQ}(t)\ &=\ \textstyle{\frac{C}{N+1}}\quad& (t>0)   \label{eq:r-fq}\\
    r_{FIFO}(t)\ & \leqslant\ \textstyle{C\cdot\frac{Cq_0}{Cq_0+NB_0}}\quad& (t>0) \label{eq:r-fifo}\\
    r_{CBQ}(t)\ &=\ \textstyle{\frac{C}{2}} \quad& (t>0) \label{eq:r-drr}\\
    r_{\name}(t)\ &=\ \textstyle{\max\left(\frac{C}{2}\cdot 2^{-\lambda t},\frac{C}{N+1}\right)} \quad& (t>0) \label{eq:r-confucius}
\end{align}
\end{subequations}
where for FIFO, $B_0$ is the initial burst size of these new flows (\eg the initial congestion window in TCP). 
We then solve for the performance degradation of the \elephant flow, $q^{max}_{P}$.
For FCT, since FQ provides the `fairest' bandwidth allocation (representing one extreme of the per-flow fairness), we use the FCT for Web flows under FQ, $T_{FQ}$ to normalize and calculate $T_{P}-T_{FQ}$ as the degree to which policy $P$ degrades Web flow performance relative to FQ.
Below we analyze four schedulers in detail.

\subsection{Fair Queueing (FQ)}
\label{app:theory-fq}
Substituting Eq.~\ref{eq:r-fq} into Eq.~\ref{eq:model}, and taking the derivatives, we have:
\begin{equation}
    \small
    \frac{{\rm d^2}}{{\rm d}t^2}s(t) + k\cdot s(t-\tau) = k\frac{C}{N+1}
\end{equation}
With loss of generality, we assume $s(\tau)=C$, meaning that before $N$ flows join, the sending rate has converged to the link capacity.
Note that the measurement loop is usually much smaller than the control loop, i.e. $\tau\ll 1/k$, we then solve the differential equation above as:
\begin{equation}
    \small
    s(t) = \left(1-\frac{1}{N+1}\right)\cos\left(\sqrt{k}(t-\tau)\right)+\frac{1}{N+1}C\quad (t>\tau)
\end{equation}
Since we are considering the transient conditions with a small $t$, where $t$ is less than the first time of $s(t) = r(t)$, we approximate the formula above with Taylor's expression:
\begin{equation}
    \small
    s(t) = C-C\frac{N}{N+1}\cdot\frac{k}{2}\cdot(t-\tau)^2\quad (t>\tau)
\end{equation}
Combine with \cref{eq:delay}, we have
\begin{equation}
    \small
    q(t)=N\left(q_0+\tau-\frac{N}{6k(N+1)}(t-\tau)^2\right)
\end{equation}
We then have the maximum queue delay as:
\begin{equation}
    \small
    \label{eq:qdelay-fq}
    q^{max}_{FQ} \geqslant q\left(\tau+\sqrt{2}{k}\right) =  N\left(\frac{2}{3}\sqrt{\frac{2}{k}}+q_0+\tau\right)
\end{equation}
As $N$ increases, $q^{max}_{FIFO}$ will also increase.

Meanwhile, by substituting the available bandwidth in \cref{eq:fct-all} with \cref{eq:r-fq}, we have $T_{FQ}$:
\begin{equation}
    \small
    T_{FQ} = \left(1+\frac{1}{N}\right)\cdot \frac{NB}{C}
\end{equation}

\subsection{FIFO}
\label{app:theory-fifo}
Since the share of available bandwidth is proportional to the share of buffer occupancy, we estimate $r_{FIFO}(t)$ as in \cref{eq:r-fifo}.
Similar to FQ, we can get:
\begin{equation}
    \small
    q(t) \geqslant \frac{1}{C}\left(\frac{NB}{q_0C}\right)\left(q_0C + \int_0^ts(t') {\rm d}t' - tC\frac{1}{\frac{NB}{q_0C}+1}\right)
\end{equation}
and then
\begin{equation}
    \small
    q^{max}_{FIFO} \geqslant q\left(\tau + \sqrt{\frac{2}{k}}\right)
\end{equation}
Consequently
\begin{equation}
    \small
    q^{max}_{FIFO} \geqslant \left(\frac{NB_0}{q_0C}+1\right)\left(\frac{2}{3}\sqrt{\frac{2}{k}}+q_0+\tau\right)
\end{equation}

\subsection{DRR}
\label{app:theory-drr}
As we can see from \cref{eq:r-drr}, the $r_{DRR}(t)$ is a special case of $r_{FQ}(t)$ with $N=1$.
Therefore, according to the delay degradation result in Eq.~\ref{eq:qdelay-fq}, we have:
\begin{equation}
    \small
    q^{max}_{DRR} \geqslant \frac{2}{3}\sqrt{\frac{2}{k}}+q_0+\tau
\end{equation}

The FCT satisfies:
\begin{equation}
    \small
    T_{DRR} = \frac{2NB}{C}
\end{equation}
In this case, 
\[
    \small
T_{DRR}-T_{FQ} =  \frac{(N-1)B}{C}
\]
diverges with $N$ and $B$.

\subsection{\name}
\label{app:theory-confucius}

For \name, we have:
\begin{equation}
    \small
    r_{\name}(t) = \frac{C}{2}e^{-\lambda t}\quad (t>0)
\end{equation}
we could then solve out (using Laplacian transform, and solve with undetermined coefficients):
\begin{equation}
    \small
    s(t) = Ae^{-\lambda (t-\tau)} + B\cos\sqrt{k}(t-\tau)
\end{equation}
where
\begin{eqnarray}
    \small
    A = & C\cdot\frac{k}{2}\cdot\frac{1}{\lambda^2+k\cdot e^{\lambda\tau}} \\
    B = & C - A
\end{eqnarray}

Still using Taylor's approximation:
\begin{equation}
    \small
\begin{array}{cl}
   s(t) &= A\left(1-\lambda(t-\tau)\right) + B\left(1-\frac{1}{2}k(t-\tau)^2\right)\\
     &= -\frac{B}{2}k(t-\tau)^2-\lambda A(t-\tau) + A+B
\end{array}
\end{equation}
Denote the root of $s(t)=0$ on $t>\tau$ as $t_0+\tau$ ($t_0>0$), we then have
\begin{equation}
    \small
q(t_0+\tau)=2e^{\lambda (t_0+\tau)}\left(q_0+\tau-\left(t_0 - \frac{\lambda A}{2C}t_0 - \frac{kB}{6C}t_0^3\right)\right)
\end{equation}
where $t_0$ satisfies:
\begin{equation}
    \small
    \label{eq:root}
    t_0 = \frac{-\lambda A+\sqrt{(\lambda A)^2 + 2Bk(A+B)}}{Bk}
\end{equation}
Thus, we have a bound of $q^{max}_{\name}$:
\begin{equation}
    \small
    q^{max}_{\name} \approx q(t_0+\tau) = f(\lambda;k,\tau,q_0)
\end{equation}
independent of $B$ or $N$. bounded. 
We expand the series as:
\begin{equation}
    \label{eq:series-qmax}
    \small
    \begin{array}{rl}
        f(\lambda) &= F_0 + F_1\lambda + F_2\lambda^2 + o(\lambda^2)  \\
        F_0 &= 2q_0 + 6\tau + \frac{8}{2\sqrt{k}}\\
        F_1 &= \frac{10}{3k} + 2q_0\tau + 2\tau^2 + \frac{4q_0}{\sqrt{k}} + \frac{16\tau}{3\sqrt{k}}\\
        F_2 &= \frac{4q_0}{k} + \frac{6\tau}{k} + q_0\tau^2 + \tau^3 + \frac{6q_o\tau}{\sqrt{k}} + \frac{11\tau^2}{\sqrt{k}}
    \end{array}
\end{equation}
Given that $\frac{1}{k}\ll q_0,\tau$, we can simplify and upper bound them into:
\begin{equation}
    \label{eq:simplified-qmax}
    \small
    q^{max}_{\name} \leqslant 6q_0 + 15\tau + \frac{8\lambda}{k} + \frac{(10q_0 + 15\tau)\lambda^2}{k}
\end{equation}

The FCT difference over the fair share for new flows is also bounded compared to other baselines.
The FCT of $N$ flows with $B$ bytes, $T$ for each flow basically follows:

Recall that $r(t) = \max(C-\frac{C}{2}2^{-\lambda t}, \frac{N}{N+1}C)$, we thus have
\begin{equation}
    \small
    T_{\name} =     \frac{(N+1)B}{C}+\frac{1}{\lambda}\cdot\left(\frac{1}{2}-\frac{1}{N}\log_2\frac{N+1}{2}-\frac{1}{2N}\right)
\end{equation}
where $t\geqslant\frac{1}{\lambda}\log_2\frac{N+1}{2}$. In this case,
\begin{equation}
    \small
    T_{\name}-T_{FQ} \leqslant \frac{1}{\lambda}\cdot\left(\frac{1}{2}-\frac{1}{N}\log_2\frac{N+1}{2}-\frac{1}{2N}\right) \leqslant \frac{\log_2 e}{\lambda}
\end{equation}

\begin{figure}
    \centering
    \subfigure[Changing $k$]{
        \includegraphics[height=2.3cm]{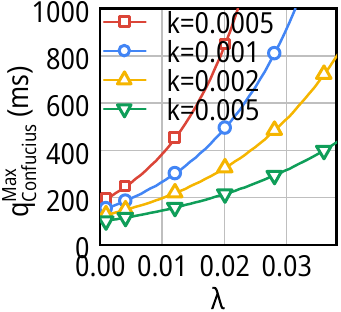}
        \label{fig:theory-kdata}
    }
    \subfigure[Changing $\tau$]{
        \includegraphics[height=2.3cm]{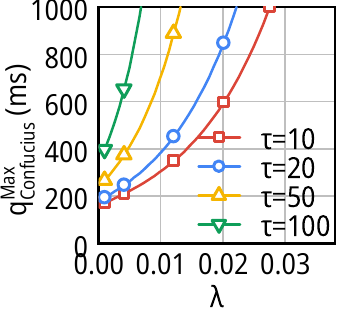}
        \label{fig:theory-tdata}
    }
    \subfigure[Changing $q_0$]{
        \includegraphics[height=2.3cm]{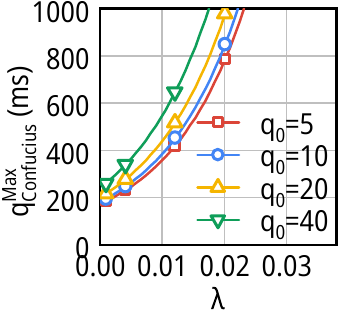}
        \label{fig:theory-q0data}
    }
    \caption{The theoretical estimation from \name under different parameter settings.}
    \label{fig:theory-plot}
\end{figure}

We further plot the unsimplified bound in different $k$ and other parameter settings in \cref{fig:theory-plot}.
Remember that the theoretical bounds are much greater than the actual experiment results, as shown in \cref{sec:eva-benchmark}.

\subsection{Responsiveness for CCAs}
\label{app:theory-responsive}

For different CCAs, we can fit their responsiveness $k$ based on their probing period in the steady state. 
From the differential equations in Eq.~\ref{eq:model} and Eq.~\ref{eq:delay}, during the steady state where $r(t)\equiv C$, we can solve that the sending rate $s(t)$ follows:
\begin{equation}
    \small
    \label{eq:steady-s}
    s(t) = C + A\cos(\sqrt{k}t + \varphi)
\end{equation}
where $A$ and $\varphi$ are undetermined coefficients.
In this case, we can know that the probing period of a CCA is $\frac{2\pi}{\sqrt{k}}$.
From the respective design of CCAs, the probing period for Copa is 5 RTTs, and for BBR is 8 RTTs.
For example, when RTT is 40 ms, we will have $k_{Copa}=0.001~(ms^{-2}$), $k_{BBR}=0.0004~(ms^{-2}$).

\section{Supplementary Experiments}
\label{app:eva}

We further evaluate the performance of \name in a series of microbenchmarking settings.

\begin{figure*}
    \centering
    \subfigure[When the RT flow uses Copa.]{
        \includegraphics[height=3.1cm]{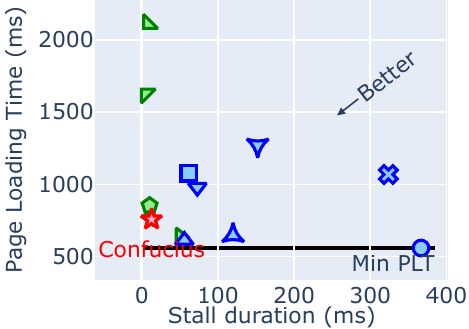}
        \label{fig:eva-tradeoff-copa-xu4g}
    }
    \subfigure[When the RT flow uses GCC.]{
        \includegraphics[height=3.1cm]{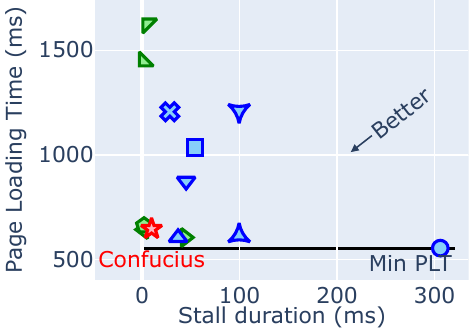}
        \label{fig:eva-tradeoff-gcc-xu4g}
    }
    \subfigure[When the RT flow uses BBR.]{
        \includegraphics[height=3.1cm]{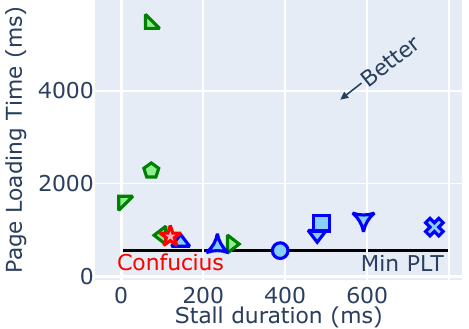}
        \label{fig:eva-tradeoff-bbr-xu4g}
    }
    \subfigure{\includegraphics[height=3cm]{figures/eva-trace-legend.pdf}}
    \caption{The trade-off between the real-time (RT) flow (stall duration) and Web flows (page loading time) on bandwidth traces \texttt{C2} (4G). We mark baselines in green if they rely on labels from end hosts, and in blue if not.}
    \label{fig:eva-tradeoff-real-xu4g}
\end{figure*}

\begin{figure*}
    \centering
    \subfigure[When the RT flow uses Copa.]{
        \includegraphics[height=3.1cm]{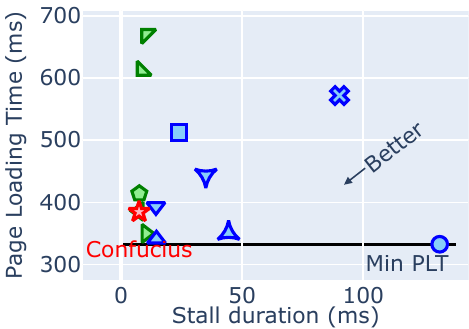}
        \label{fig:eva-tradeoff-copa-xu5g}
    }
    \subfigure[When the RT flow uses GCC.]{
        \includegraphics[height=3.1cm]{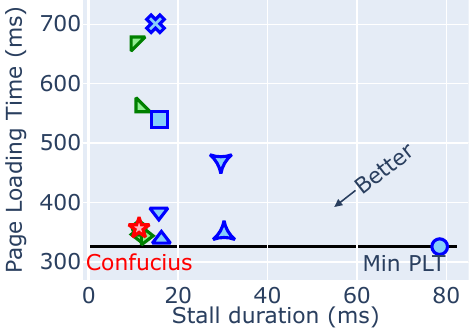}
        \label{fig:eva-tradeoff-gcc-xu5g}
    }
    \subfigure[When the RT flow uses BBR.]{
        \includegraphics[height=3.1cm]{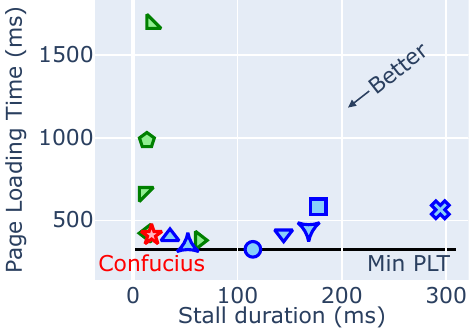}
        \label{fig:eva-tradeoff-bbr-xu5g}
    }
    \subfigure{\includegraphics[height=3cm]{figures/eva-trace-legend.pdf}}
    \caption{The trade-off between the real-time (RT) flow (stall duration) and Web flows (page loading time) on bandwidth traces \texttt{C3} (5G). We mark baselines in green if they rely on labels from end hosts, and in blue if not.}
    \label{fig:eva-tradeoff-real-xu5g}
\end{figure*}

\begin{figure*}
    \centering
    \subfigure[When the RT flow uses Copa.]{
        \includegraphics[height=3.1cm]{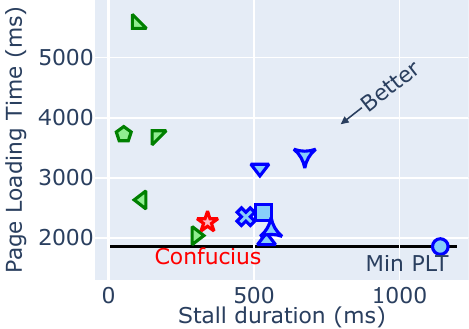}
        \label{fig:eva-tradeoff-copa-officewifi}
    }
    \subfigure[When the RT flow uses GCC.]{
        \includegraphics[height=3.1cm]{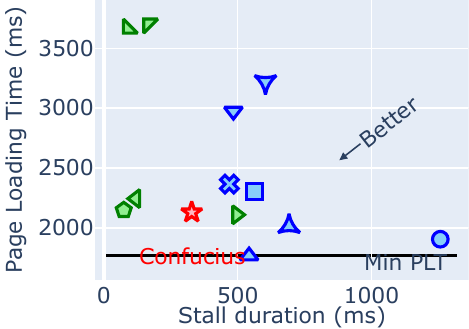}
        \label{fig:eva-tradeoff-gcc-officewifi}
    }
    \subfigure[When the RT flow uses BBR.]{
        \includegraphics[height=3.1cm]{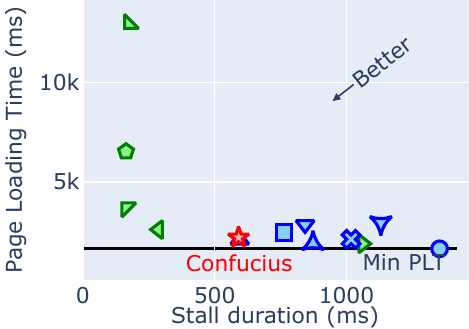}
        \label{fig:eva-tradeoff-bbr-officewifi}
    }
    \subfigure{\includegraphics[height=3cm]{figures/eva-trace-legend.pdf}}
    \caption{The trade-off between the real-time (RT) flow (stall duration) and Web flows (page loading time) on bandwidth traces \texttt{W1} (Office WiFi). We mark baselines in green if they rely on labels from end hosts, and in blue if not.}
    \label{fig:eva-tradeoff-real-officewifi}
\end{figure*}

\begin{figure*}
    \centering
    \subfigure[When the RT flow uses Copa.]{
        \includegraphics[height=3.1cm]{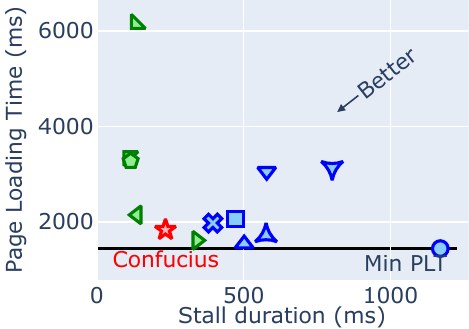}
        \label{fig:eva-tradeoff-copa-restwifi}
    }
    \subfigure[When the RT flow uses GCC.]{
        \includegraphics[height=3.1cm]{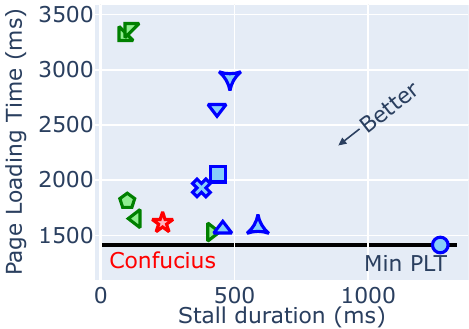}
        \label{fig:eva-tradeoff-gcc-restwifi}
    }
    \subfigure[When the RT flow uses BBR.]{
        \includegraphics[height=3.1cm]{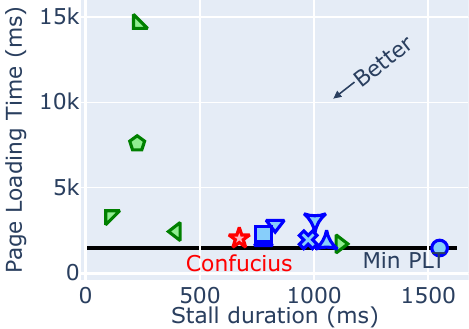}
        \label{fig:eva-tradeoff-bbr-restwifi}
    }
    \subfigure{\includegraphics[height=3cm]{figures/eva-trace-legend.pdf}}
    \caption{The trade-off between the real-time (RT) flow (stall duration) and Web flows (page loading time) on bandwidth traces \texttt{W2} (Restaurant WiFi). We mark baselines in green if they rely on labels from end hosts, and in blue if not.}
    \label{fig:eva-tradeoff-real-restwifi}
\end{figure*}

\subsection{Results for other traces}
\label{app:eva-traces}

\begin{figure}
    \centering
        \includegraphics[height=2.8cm]{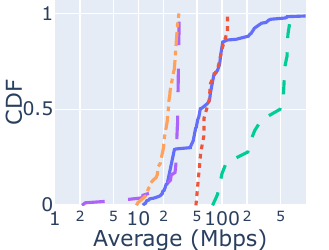}
        \includegraphics[height=2.8cm]{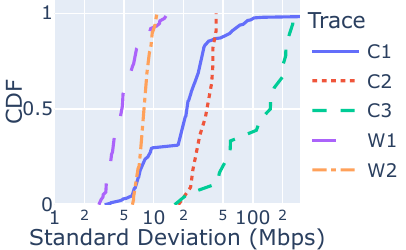}
    \caption{The bandwidth distribution of five trace datasets we use in this paper. X-axes are log-scaled.}
    \label{fig:eva-bwstat}
\end{figure}

We further present the results of \name over other bandwidth trace datasets (C2: Cellular 4G; C3: Cellular 5G; W1: Office WiFi; W2: Restaurant WiFi) in \cref{fig:eva-tradeoff-real-xu4g,fig:eva-tradeoff-real-xu5g,fig:eva-tradeoff-real-officewifi,fig:eva-tradeoff-real-restwifi}.
The experiment setting follows the one in \cref{fig:eva-tradeoff-real}.
The average and standard deviation of bandwidth of these traces are presented in \cref{fig:eva-bwstat}.
We can clearly see that \name always pushes the Pareto-optimal frontier of baselines not requiring labels (blue baselines) front.
This demonstrates the robustness of \name across different bandwidth datasets.

\subsection{Fairness}
\label{app:eva-jfi}
\begin{figure}
    \centering
    \includegraphics[height=2.7cm]{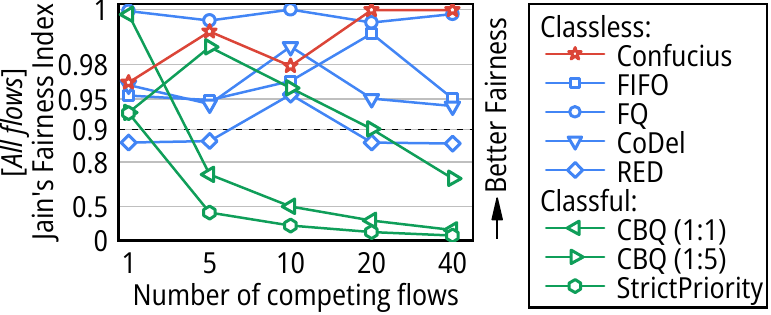}
    \caption{Jain's fairness index (JFI) as the workload changes. The fairness of classful solutions (\eg CBQ) is heavily sensitive to workload variations. For instance, CBQ with different weights (1:1 or 1:5) will result in poor fairness (JFI$<$0.9) in certain workloads. Y axis is not lin-scaled.}
    \label{fig:motiv-param}
\end{figure}

Classful schemes such as CBQ, which splits packets into classful queues of configurable service rate or strict priority, which only dequeues packets of lower priority if high priority is empty, protect the real-time flow. 
However, classful schemes also result in unfair allocations because they overpenalize (or even starve) web traffic which experiences high page load times (PLTs) as shown in \cref{fig:design-weight-drr}.
While, in theory, CBQ could be configured to be fair, that requires knowledge of the exact workload (ratio of flows between classes) over very short time intervals, which is in practice infeasible. 
For example, we measure the fairness that different schedulers can provide while changing the number of competing flows to the real-time flow in \cref{fig:motiv-param}.
Modifying CBQ's configuration improves JFI for a subset of the workloads: CBQ (1:1) works well when there are two flows competing while CBQ (1:5) achieves a good JFI when there are five competing flows -- they both degrade as the number of flows changes.
Even we change the ratio, such a phenomenon still exists -- the JFI of CBQ (1:5) is the highest when there are 5 new flows, but degrades drastically in other workloads.
In contrast, \name can relatively keep the JFI consistent across different number of competitors.

\begin{figure}
    \centering
    \begin{minipage}{.48\linewidth}
        \centering
        \includegraphics[height=2.4cm]{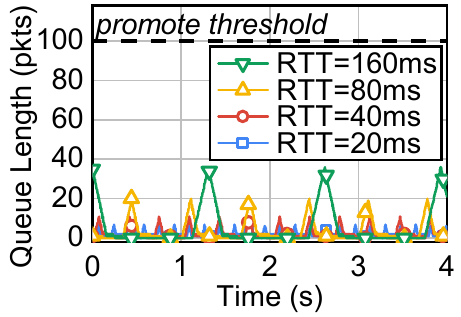}
        \caption{The hysteresis design in \name (\cref{sec:classify-hysteresis}) is able to absorb the fluctuations caused by probing from CCAs.}
        \label{fig:eva-probing-timeline}
    \end{minipage}
    \hfill
    \begin{minipage}{.48\linewidth}
        \centering
        \includegraphics[height=2.4cm]{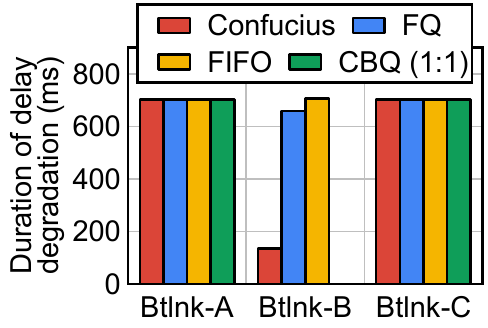}
        \caption{When the bottleneck is elsewhere, \name maintains the same performance as existing mechanisms.}
        \label{fig:eva-moving-btlbw}
    \end{minipage}
\end{figure}

\subsection{Working with Bandwidth Probing}
\label{app:eva-probing}

Some recent CCAs proposed to periodically probe the available bandwidth by overshooting the network, which might introduce noises in classifying the buffer occupancy of flows in \name.
Some recent examples for video streaming include Sprout~\cite{nsdi2013sprout}, PCC (probing up to 5\%)~\cite{nsdi2015pcc}, and BBR (probing 25\%)~\cite{queue2016bbr}.
We evaluate how \name is able to handle the bandwidth probing from CCAs.
We first run one BBR flow, which is the most aggressive one among these bandwidth probing CCAs, and change the RTT from 20 ms to 160 ms since the probing period is counted in the unit of RTT.
As shown in \cref{fig:eva-probing-timeline}, with the other settings the same as \cref{fig:exp-setup}, the queue fluctuations never go across the threshold of reclassification of the flow.
This is due to the hysteresis design in \cref{sec:classify-hysteresis} -- \name deliberately makes conservative decisions in the classification of flows to smoothize the noises out.
This can also be validated from \cref{fig:classify-results}: the classification results are stable all the time even if BBR periodically probes the bandwidth.
Therefore, \name is able to work well with bandwidth-probing CCAs.

\subsection{Working with Different Bottleneck}
\label{app:eva-btlbw}

\begin{figure}
    \centering
    \includegraphics[width=.9\linewidth]{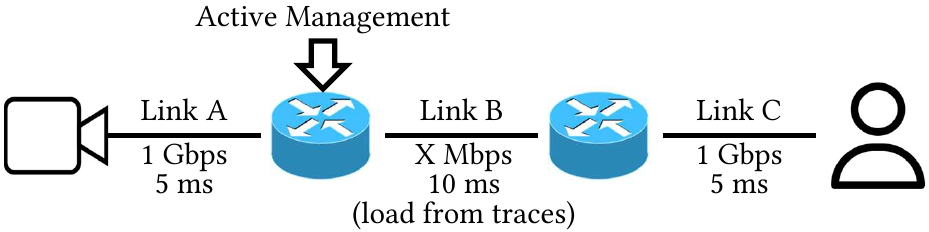}
    \caption{Experimental setup for multiple bottlenecks.}
    \label{fig:exp-setup}
\end{figure}

We further evaluate the end-to-end performance when the bottleneck is not where \name is deployed.
\name is able to reduce the latency volatility when it is deployed on the bottleneck router.
Our further experiments show that \name does not introduce side effects when the bottleneck is before or after the router deployed with \name.
We still deploy queue management mechanisms to the router before link B and respectively rate-limit the link A, B, and C in \cref{fig:exp-setup} to 20 Mbps:
\begin{itemize}
    \item \texttt{Btlnk-A}. When link A is limited while the other two links are set to 100 Mbps, the bottleneck is before the place of \name.
    \item \texttt{Btlnk-B}. The case when link B is limited is what we mainly evaluated in this section, where \name is at the bottleneck.
    \item \texttt{Btlnk-C}. When link C is limited, the bottleneck is after the place of \name.
\end{itemize}
For those unmanaged routers, they adopt FIFO as their default mechanism.
As shown in \cref{fig:eva-moving-btlbw}, the performance is only affected by the mechanism deployed at the bottleneck.
When \name is not at the bottleneck (e.g., link A or C), the performance is the same no matter what mechanism is deployed at link B.
It is worth to note that as discussed in a series of papers~\cite{sigcomm2022zhuge, conext2015qprobe}, the last-mile routers (e.g., cellular base stations, home wireless APs) are the bottleneck for most of the congestions, in which case deploying \name will achieve significant performance benefits.

\begin{figure}
    \centering
    \subfigure[Stall duration of the real-time (existing) flow.]{
        \includegraphics[height=2.4cm]{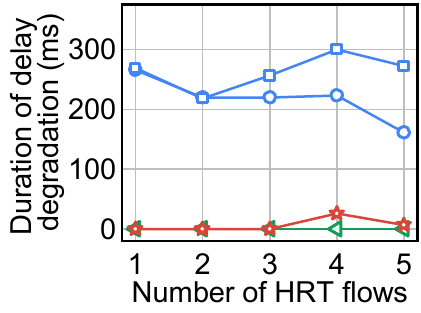}
        \label{fig:multivideo-delay}
    }\hfill
    \subfigure[PLT of Web (new) flows.]{
        \includegraphics[height=2.4cm]{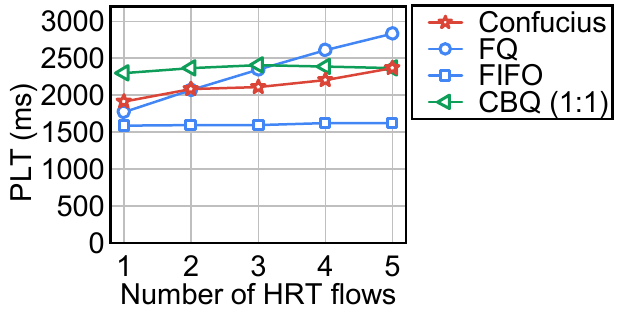}
        \label{fig:multivideo-plt}
    }
    \caption{We increase the number of simultaneous real-time flows, and measure the results again with the Alexa dataset.}
    \label{fig:eva-multivideo}
\end{figure}

\subsection{Multiple Real-time Flows Competition}
\label{app:eva-multivideo}

We further evaluate the performance when there are multiple real-time flows running simultaneously.
We reproduce the experiments in \cref{fig:eva-tradeoff-copa} but change the number of real-time flows from 1 to 5.
The average duration of delay degradation of real-time flows, and the PLT of Web flows are presented in \cref{fig:eva-multivideo}.
\name is able to provide a consistent performance for multiple real-time flows in the same time -- the delay degradation is consistently negligible independent of the number of concurrent real-time flows and the PLT stays roughly the same place compared to the baselines.
Note that since \name is designed for last-mile routers, 5 concurrent flows should be able to cover most scenarios~\cite{sigcomm2022zhuge}.

\section{Related Work}
\label{app:related}

\parahead{Queue management solutions.}
There are numerous efforts on queue management for routers.
Besides the solutions we introduced in \cref{sec:2} and \cref{sec:eva-setup}, there are even more AQMs proposed back to 2000s~\cite{infocom2001sfb, globecom2002green, comnet2005yellow, icc2003black, sigcomm2003afd}.
As we discussed in \cref{sec:2}, these AQMs cannot meet the requirement of providing consistent performance and fairness during transient events.
At the same time, recent delay- or rate-based CCAs, which are commonly used in real-time flows, are not responsive to such dropping-based or ECN marking-based AQMs.
Further, datacenter flow scheduling schemes~\cite{nsdi2015pias, sigcomm2013pfabric} or buffer management \cite{sigcomm2022abm} are designed for homogeneous flows (sometimes with labelled packets) and are not suitable for heterogeneous flows in home routers in the wide-area network.

\parahead{Optimizations for latency consistency.} 
Multiple schemes aim at offering consistent low latency for latency-sensitive applications such as videoconferencing  either at the end hosts~\cite{nsdi2018salsify, ton2017webrtc, nsdi2018copa}, and/or in-network~\cite{sigcomm2022zhuge, nsdi2020abc}.
Besides, there are also application-specific solutions such as frame-rate or bit-rate adaption~\cite{nsdi2023afr, nsdi2018salsify} and latency compensation~\cite{mmsys2020latency}.
\name is orthogonal to such solutions.

\parahead{Inter-flow fairness.}
The fairness across flows dates back to the birth of congestion control~\cite{sigcomm1988congestion}.
Recent work analyzes fairness in different scenarios~\cite{imc2021revisiting} or defines fairness with different applications~\cite{hotnets2021macc, hotnets2019harm}.
There are also measurements investigating the inter-CCA fairness with emerging CCAs~\cite{cloudnet2018tcp, molnar2009comprehensive, bbr2017icnp}.
Instead, \name is also able to maintain the long-term fairness across flows.

\end{document}